\documentclass{aa}  

\usepackage{graphicx}
\usepackage{txfonts}
\usepackage{hyperref}
\usepackage{float}
\begin{document}

   \title{Protostellar Outflows at the EarliesT Stages (POETS) \\
    VI. Evidence of disk-wind in G11.92-0.61 MM1}

\titlerunning{POETS VI. Disk-wind traced by molecular lines in G11.92-0.61 MM1}

   \author{O. S. Bayandina\inst{1}
          \and
          L. Moscadelli\inst{1}
          \and 
          R. Cesaroni\inst{1} 
          \and
          M. T. Beltrán\inst{1}
          \and
          A. Sanna\inst{2}
          \and
          C. Goddi\inst{2}
          }

             \institute{INAF – Osservatorio Astrofisico di Arcetri, Largo E. Fermi 5, 50125 Firenze, Italy\\
             \email{olga.bayandina@inaf.it}
             \and
             INAF – Osservatorio Astronomico di Cagliari, Via della Scienza 5, 09047 Selargius (CA), Italy 
             }

   \date{Received ; accepted }

 
  \abstract
   {Magnetohydrodynamic disk-winds are thought to play a key role in the formation of massive stars by providing the fine-tuning between accretion and ejection, where excess angular momentum is redirected away from the disk, allowing further mass growth of a young protostar. However, only a limited number of disk-wind sources have been detected to date.\ To better constrain the exact mechanism of this phenomenon, expanding the sample is critical.}
   {We performed a detailed analysis of the disk-wind candidate G11.92-0.61 MM1 by estimating the physical parameters of the massive protostellar system and constraining the wind-launching mechanism.
   }
   {Atacama Large Millimeter/submillimeter Array (ALMA) Band 6 observations of G11.92-0.61 MM1 were conducted in September 2021 with ALMA's longest baselines, which provided a synthesised beam of $\sim$30 mas. We obtained high-resolution images of the CH$_3$CN ($\varv_8=1$ and $\varv=0$), CH$_3$OH, SO$_2$, and SO molecular lines, as well as the 1.3 mm continuum.}
   {Our high-resolution molecular data allowed us to refine the parameters of the disk-outflow system in MM1. The rotating disk is resolved into two regions with distinct kinematics: the inner region ($<$300 au) is traced by high-velocity emission of high-excitation CH$_3$CN lines and shows a Keplerian rotation; the outer region ($>$300 au), traced by mid-velocity CH$_3$CN emission, rotates in a sub-Keplerian regime.  
   The central source is estimated to be $\sim$20 $M_{\odot}$, which is about half the mass estimated in previous lower-resolution studies.
   A strong collimated outflow is traced by SO and SO$_2$ emission up to $\sim$3400 au around MM1a. 
   The SO and SO$_2$ emissions show a rotation-dominated velocity pattern, a constant specific angular momentum, and a Keplerian profile that suggests a magneto-centrifugal disk-wind origin with launching radii of $\sim$50-100 au. 
   }
   {G11.92-0.61 MM1 appears to be one of the clearest cases of molecular line-traced disk-winds detected around massive protostars.}

   \keywords{Accretion, accretion disks - Magnetohydrodynamics (MHD) - Stars: winds, outflows - Stars: formation - Stars: massive - Stars: individual: G11.92-0.61
               }

   \maketitle
%

\section{Introduction}

Star formation occurs through the interplay of accretion and ejection -- seemingly antagonistic forces that are in fact two sides of the same coin. A surprising amount of value is hidden in the thin and often overlooked edge of the coin, the region of interaction between accretion and ejection.

As accretion feeds a protostar, ejection removes angular momentum from the accreting matter, allowing the process to continue. Although the fundamental concept is clear, formulating the actual mechanism of gas-launching requires a great deal of theoretical and observational work. 
Theoretical works have come a long way from assuming the impact of hydrodynamics and radiation only (for example, \citealt{Canto1980}), by, for example, including the role of magnetic fields via ideal magnetohydrodynamics (MHD; for example, \citealt{Blandford1982}) and adding further physics, such as non-ideal MHD effects (for example, \citealt{Suriano2019}). 

Observational studies first examined large-scale low-velocity outflows (for example, \citealt{Bally1983}) before turning to jets (compact, collimated shocked gas ejected at high speeds and  powering large-scale outflows; for example, \citealt{McCaughrean1994, Frank2014}). Modern observing facilities are now finally able to achieve the resolutions needed to study the launching region of jets.
Jets are modelled to be launched by rotation and gravitational energy along the magnetic field lines, but the precise acceleration mechanism is still debated. 
The `X-wind' theory assumes that jets are launched from the inner edge of an accretion disk, a small region within $<$1 au of the star that is dominated solely by the stellar magnetic field \citep{Shu1995}. In contrast, the `disk-wind' theory assumes that jets are launched at radii larger by two to three orders of magnitude, where the poloidal component of the disk's magnetic field dominates \citep{Pudritz2007}.

Which theory is correct, 
X-wind or disk-wind, is still an open question. Although the disk-wind theory has gained more support lately, the evidence in its favour was mostly obtained for low-mass protostars \citep{Pascucci2023}.
The main obstacle to studying the disk-wind emission in massive protostars is the difficulty in achieving a spatial resolution sufficient to access the region of disk-jet interaction, as well as in identifying suitable tracers. 
The high resolution required to resolve the jet's launching regions and obtain evidence for the existence of a disk-wind has been achieved with very long baseline interferometry observations of masers: SiO masers in Orion-BN/KL \citep{Matthews2010} and H$_2$O masers in IRAS 21078+5211 \citep{Moscadelli2022}. 
Such a high resolution has been achieved with the Atacama Large Millimeter/submillimeter Array (ALMA) for Orion Source I, which is located at a close  distance of 418 pc \citep{Kim2008}, where \cite{Hirota2017} found compelling evidence for a disk-wind through the detection of a rotating bipolar outflow traced by SiO emission. In this case, neither stellar winds nor X-winds were found to be consistent with the estimated launching radius ($>$10 au) and velocity (10 km/s).
For a few more maser sources, the possibility of the presence of disk-winds was only hypothesised (for example, \citealt{Moscadelli2011, Sanna2012}).

With the aim of shedding light on the nature and properties of protostellar winds and jets, we started the large project POETS \citep[Protostellar Outflows at the EarliesT Stages;][]{Moscadelli2016, Sanna2018, Moscadelli2019}.
Initially, the project was based on a high-resolution survey of
radio continuum and water masers, which were observed with the goal of tracing free-free emission from ionised gas in the inner outflow
cavities and fast shocked layers of outflowing gas. In recent years, we have selected a number of targets from the POETS sample for follow-up observations with multiple molecular lines in order to provide a view of the diverse phenomenology and to obtain information on outflow tracers. Here, we report the results for one of the POETS targets, G11.92-0.61.

The star-forming region G11.92-0.61 is a promising target for studying the ejection process in massive protostars.
 Located in the Galactic plane, at a distance of $3.37^{+0.39}_{-0.32}$ kpc \citep{Sato2014}, G11.92-0.61 is associated with an infrared dark cloud that is actively forming both low- and high-mass stars \citep{Cyganowski2017}. The region has been extensively studied in a variety of wavelengths, which revealed a complex and dynamic environment. 

At first, G11.92-0.61 attracted attention when it was identified as an extended green object (EGO; \citealt{Cyganowski2008}) and as such was considered to be a massive protostellar outflow candidate. The 4.5 $\mu$m image of G11.92-0.61 showed a bipolar morphology consisting of a north-east (NE) and a south-west (SW) region \citep{Cyganowski2008, Cyganowski2011}. Initial (sub-)millimetre observations of the region detected only three massive compact millimetre-continuum cores \citep[MM1-MM3;][]{Cyganowski2011}, while follow-up ALMA observations revealed 16 more low-mass millimetre-continuum cores \citep{Cyganowski2017}. The detected millimetre-continuum cores of low and high masses \citep{Cyganowski2017}, as well as the 6.7 GHz methanol \citep{Cyganowski2009} and 22 GHz water masers (for example, \citealt{Moscadelli2016}), appeared to be predominantly associated with the NE region \citep{Cyganowski2011}, with only a cluster of 44 GHz methanol masers found in the SW region \citep{Cyganowski2009, Cyganowski2011, Cyganowski2017}.  Out of the EGO sample studied in \cite{Cyganowski2008}, G11.92-0.61 is the only one that might contain multiple massive protostars, as suggested by the association with two distinct sites of 6.7 GHz methanol maser emission.

The brightest and most massive millimetre core of the region, MM1, is considered an example of a forming proto-O star with an active outflow and ongoing accretion. The central source of MM1 seems to be very young, showing weak ($\sim$1 mJy) centimetre-continuum emission but rich spectral line and maser emission \citep{Cyganowski2011b, Ilee2016, Moscadelli2016}. The Submillimeter Array (SMA) and ALMA studies of hot-core molecules in MM1, presented in \cite{Ilee2016} and \cite{Ilee2018}, respectively, revealed the presence of a massive Keplerian accretion disk with an enclosed mass of the central object  estimated to be $\sim$40 M$_\odot$  \citep{Ilee2018}. However, the detected luminosity of only $\sim$10$^4$ L$_{\rm sun}$  \citep{Cyganowski2011, Moscadelli2016} appeared to be too low for such a massive object; \cite{Ilee2018} proposed that such a discrepancy might be explained if the luminosity comes from accretion or if multiple stars are present at the disk centre.
The detection of a weak 1.3 mm continuum source to the south-east of MM1, which was denominated MM1b, was interpreted as a sign of one of the first observed examples of disk fragmentation around a high-mass protostar \citep{Ilee2018}. However, the case calls for further investigation as MM1b is separated from MM1 by $\sim$2000 au and  in fact seems to be located outside the disk.

MM1 drives a massive, collimated bipolar molecular outflow that is detected in numerous tracers \citep{Cyganowski2011}. The outflow has a position angle (PA) of $\sim$52$^{\circ}$, with the red lobe propagating to the NE and the blue lobe propagating to the SW \citep{Cyganowski2011, Ilee2016}.
One of the sites of  6.7 GHz methanol maser emission detected in the region is associated with MM1 \citep{Cyganowski2009}. However, the methanol maser does not coincide with the continuum peak, and is located to the south, downstream of the outflow. The intense 22 GHz water masers \citep{Hofner1996, Breen2011, Sato2014} associated with MM1 show a bipolar structure and velocity pattern similar to the large-scale outflow \citep{Moscadelli2016, Moscadelli2019}, but the kinematic structure traced by the water emission remains unclear. \cite{Moscadelli2019} have suggested that the 22 GHz water maser may be explained either as tracing  a disk-wind or as being due to central source multiplicity.

In this article we present the results of our higher-angular-resolution ALMA observations of the high-mass core G11.92-0.61 MM1. We aim to provide a detailed view of the physical and chemical properties of G11.92-0.61 MM1 and clarify the structure of the source.

\section{Observations and data reduction}

The ALMA observations of G11.92-0.61 MM1 were conducted on September 29, 2021 (project 2019.1.01639.S) for a total observing time of 1.5 hours. The array configuration was C43-9/10 with 45 antennas. The observations provided a synthesised beam with a size of 28 mas $\times$ 25 mas, and PA=-66$^{\circ}$ (corresponding to a linear resolution of $\sim$100 au at the distance of 3.37 kpc to the source).

Two sources, J1924-2914 and J1830-1606, were observed as flux, delay, and bandpass calibrators. The quasar J1825-1718 was used as phase calibrator. 

The spectral setup consisted of five spectral windows centred at 220.63, 222.15, 221.44, 219.91, and 219.44 GHz in Band 6. Each spectral window was sampled with 960 channels each of $\sim$0.6 MHz wide, providing a spectral resolution of $\sim$0.7 km s$^{-1}$.

Spectral line identification was performed in the Cube Analysis and Rendering Tool for Astronomy (CARTA\footnote{\url{https://cartavis.org}}; \cite{Comrie2020}) and MAdrid Data CUBe Analysis (MADCUBA; \cite{Martin2019}) 
packages. The molecular line data modelling and estimation of the physical parameters were carried out using the SLIM tool of MADCUBA.

Imaging and data cubes analysis were performed with the Common Astronomy Software Applications (CASA\footnote{\url{http://casa.nrao.edu}}; \cite{CASA2022}) package. 
The spectral line and continuum images were created using the CASA task \textit{tclean} with natural weighting.  
From the obtained line cubes, we computed Moment 0 and 1 maps using the CASA task \textit{immoments}. The velocity range was chosen to include the target line only, and the threshold was set at a 5$\sigma$ rms level. 
Additionally, we produced position–velocity (PV) diagrams for the line data using the GILDAS\footnote{\url{https://www.iram.fr/IRAMFR/GILDAS}} software.
Finally, we also created `peak maps' with the goal of determining the emission peak position in individual velocity channels. Given that emission in some channels of the line cubes was too diffuse or weak to be properly fitted, we `tapered' the visibilities to create lower resolution cubes with the synthesised beam of $\sim$0$\farcs$15 (which was found to be the optimal balance between retaining high resolution and obtaining distinct peaks).
Then we fitted the emission peaks in each channel  with the task \textit{imfit}, which yielded  position uncertainties from $\sim$0.8 mas ($\sim$3 au) for bright and compact emission and up to $\sim$8 mas ($\sim$30 au) for weak, extended emission.

\section{Results}

\subsection{1.3 mm continuum}

\begin{figure}[h]
\centering
  \includegraphics[width=90mm]{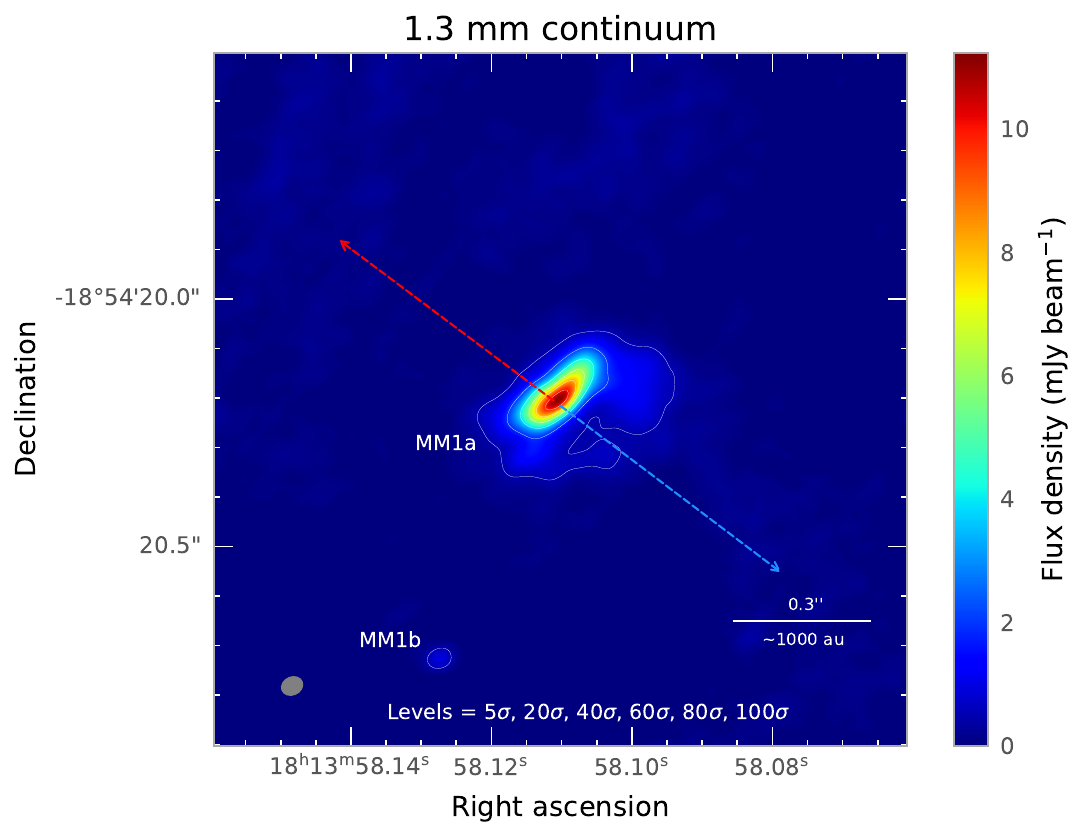}  \\
\caption{ALMA 1.3 mm continuum image of MM1a and MM1b. The contour levels are indicated at the bottom of the image with 1$\sigma$=$\sim$0.1 mJy beam$^{-1}$ (Table \ref{tab:cont}). The synthesised beam is indicated as a grey ellipse in the lower-left corner. The blue and red arrows indicate the orientation of the large-scale $^{12}$CO outflow from \cite{Ilee2016}.}
\label{fig:cont}
\end{figure}

The 1.3 mm continuum yields detection of two emission peaks corresponding to the sources MM1a and MM1b. The 1.3 mm continuum image is presented in Fig. \ref{fig:cont} and the parameters of the detected continuum sources are listed in Table \ref{tab:cont}. 
The flux density of MM1b is $\sim$1 mJy, corresponding to $\sim$5$\sigma$ level, and thus the detection is marginal.
The 1.3 mm continuum emission of MM1a is found to come from an NW-SE elongated region of $\sim$0$\farcs$16 ($\sim$500 au) with PA $\sim$130$^{\circ}$ and integrated flux density of $\sim$58 mJy (Table \ref{tab:cont}). 
Interestingly, Fig. \ref{fig:cont} shows some weaker emission towards the south and west from the peak, apparently surrounding the axis of the blue lobe of the outflow observed by \cite{Ilee2016}.

Considering the measured flux density at 1.3 mm and assuming that the dust is optically thin and in thermal equilibrium with the gas at average temperature of $\sim$300 K (for CH$_3$CN and CH$_3$OH; see Sect. \ref{ResLines} and Table \ref{tab:l-param}), we can use the standard approach of \cite{Hildebrand1983}. More precisely, by applying Eqs. (1) and (2)  from \cite{Schuller2009} to estimate the total gas mass and column density of H$_2$, we derive $M$ $\sim$2.5 $M_{\odot}$ and $N_{\rm H_2}$ = $\sim$1$\times$10$^{23}$ cm$^{-2}$. These numbers are close to the ones obtained in the lower-resolution 1.3 mm continuum observations of \cite{Ilee2018}. However, we note that our estimates may be lower limits because part of the continuum emission could have been resolved out; additionally, the column density estimate is probably a lower limit because the continuum source is not resolved in the NE-SW direction.

\begin{table*}
\caption{Parameters of the detected continuum peaks.}
\label{tab:cont}
\centering  
\begin{tabular}{lccccccc}
\hline\hline   
  \multicolumn{1}{c}{Source} &
RA(J2000) & Dec(J2000)  & Size & PA &
Int. flux &
Peak flux  \\ 
 &
($^h$~$^m$~$^s$) & ($^\circ$~$\arcmin$~$\arcsec$)  & ($\arcsec$$\times$$\arcsec$) & ($^\circ$) &
(mJy) & 
(mJy/beam)  \\
\hline
  MM1a & 18:13:58.1106$\pm$0.0001  & -18:54:20.199$\pm$0.001 & 0.16$\times$0.05 & 132 & 57.8$\pm$1.4 &  10.3$\pm$0.2 \\
 MM1b\tablefootmark{(a)} & 18:13:58.1274$\pm$0.0002 & -18:54:20.726$\pm$0.002 & - & - & 1.0$\pm$0.2 & 1.0$\pm$0.1 \\
\hline                                  
\end{tabular}
\tablefoot{$^{(a)}$Tentative detection at 5$\sigma$ level (the rms of the continuum image is $\sim$0.1 mJy beam$^{-1}$).}
\end{table*}

\subsection{Molecular lines} \label{ResLines}

\begin{figure*}
\centering
\begin{tabular}{ccc}
  \includegraphics[width=60mm]{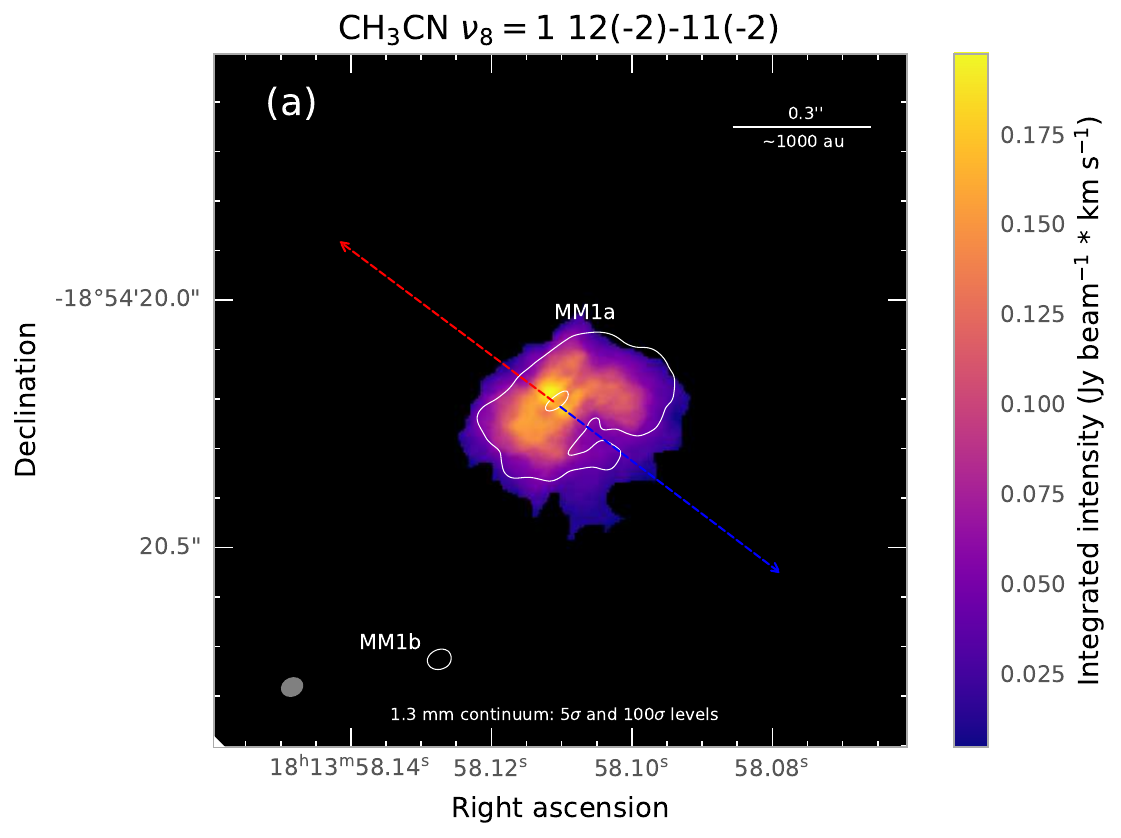}  &
  \includegraphics[width=60mm]{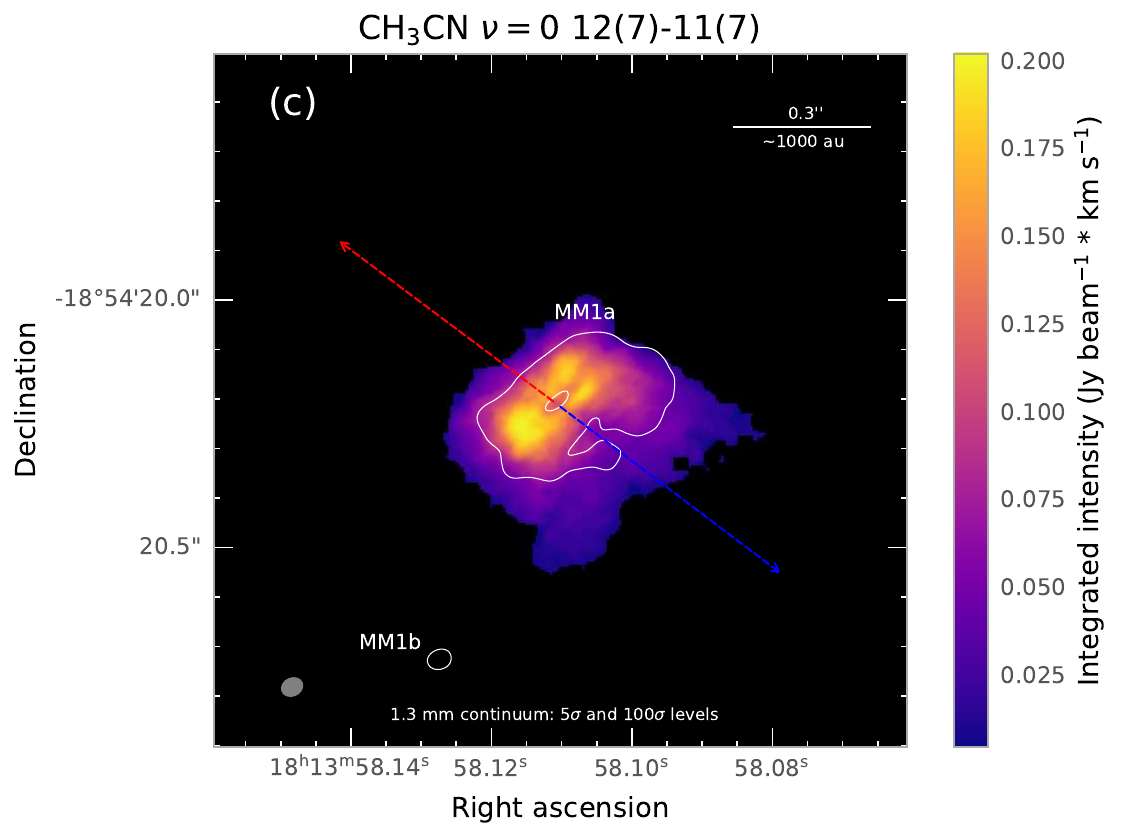}  &
  \includegraphics[width=60mm]{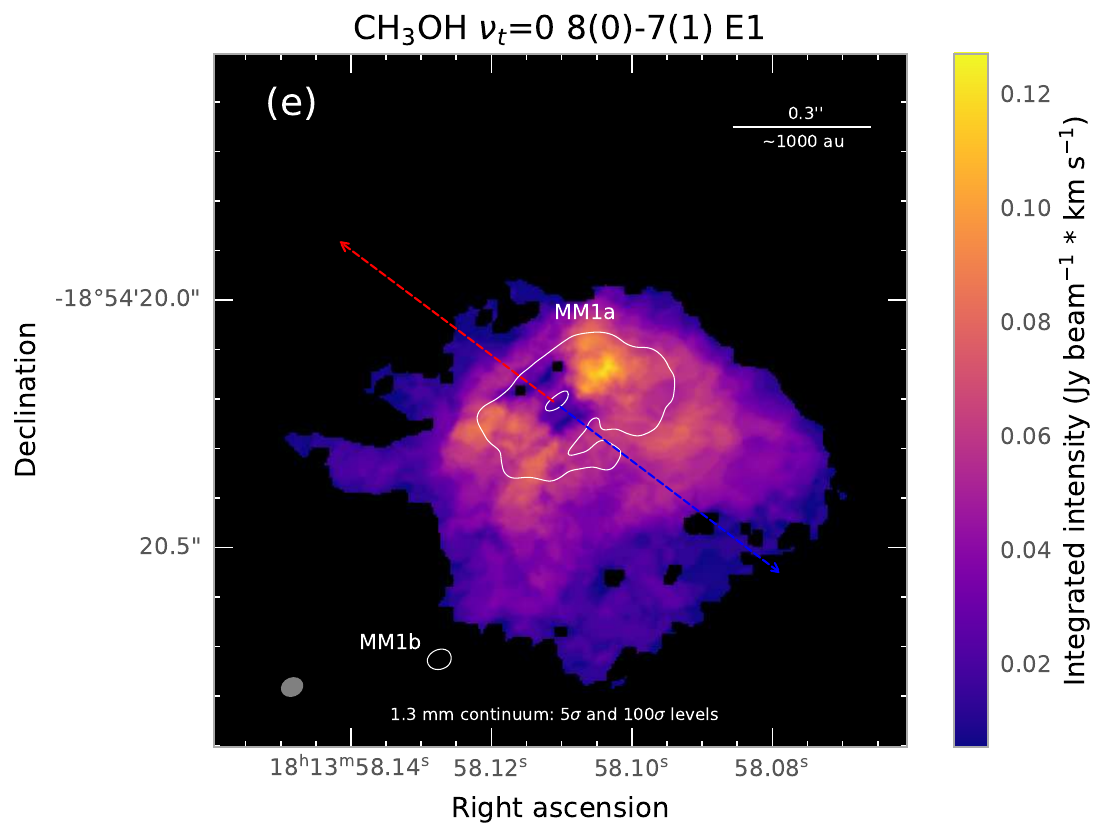}  \\
   \includegraphics[width=60mm]{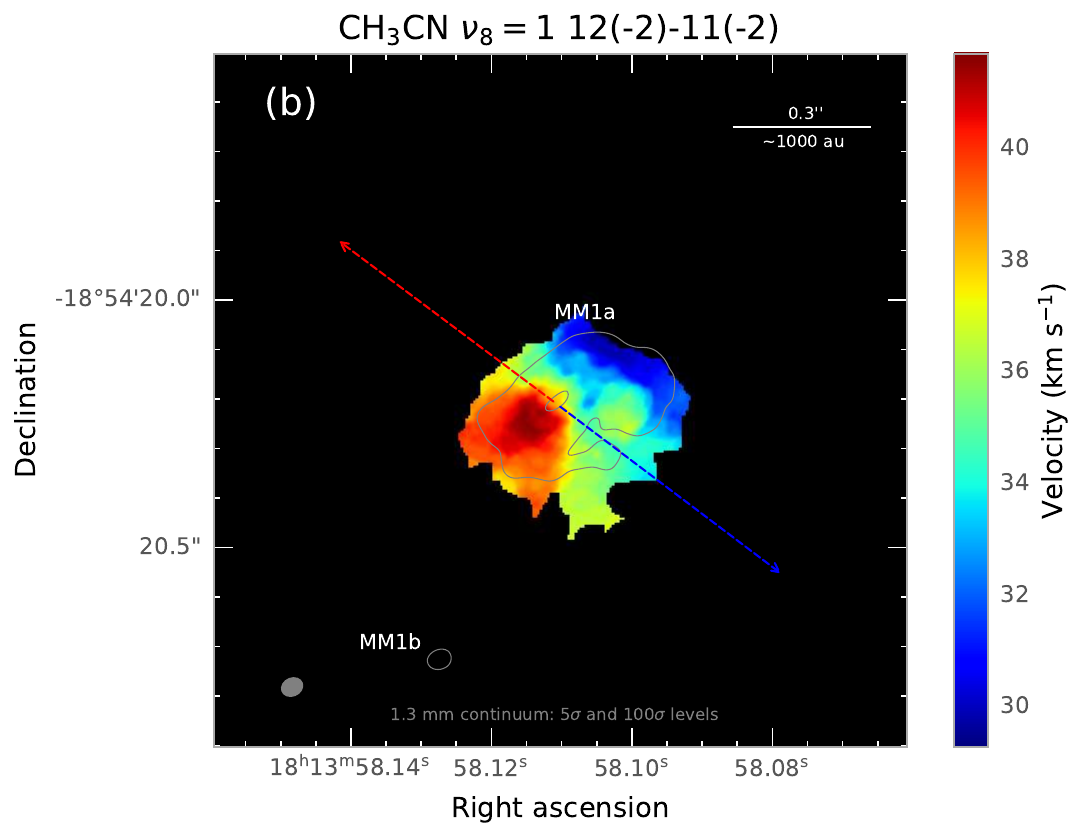}  &
  \includegraphics[width=60mm]{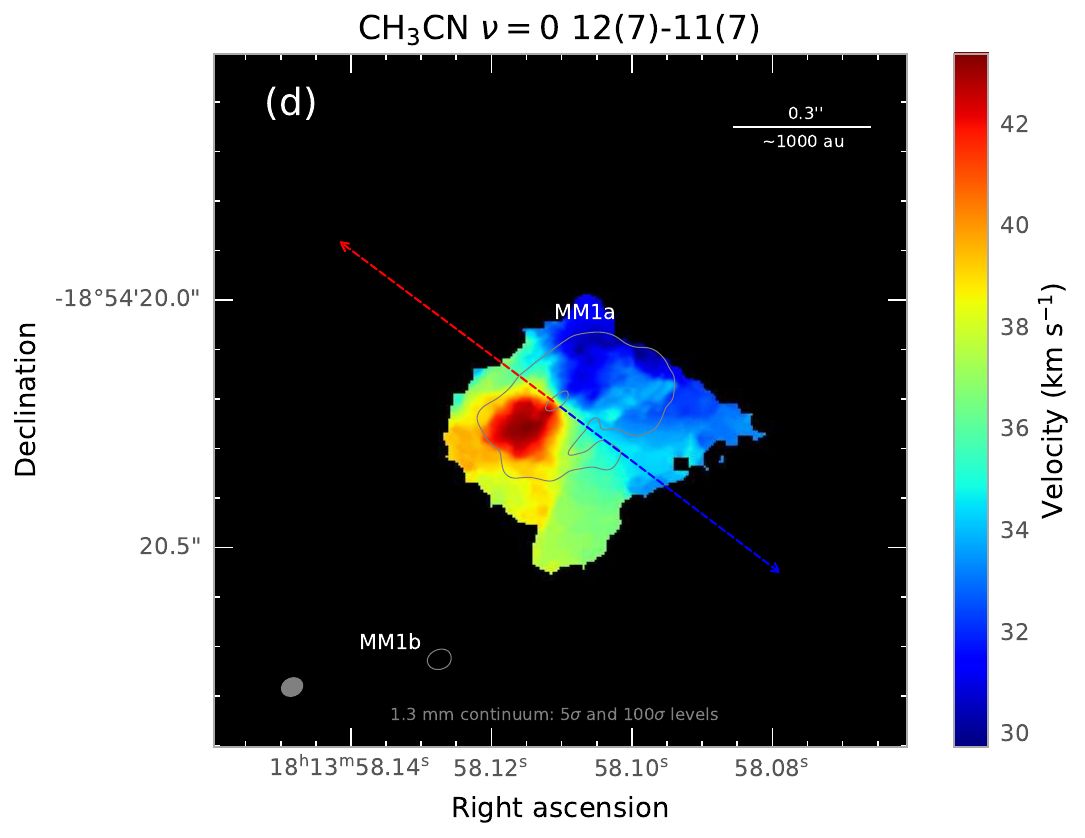}  &
  \includegraphics[width=60mm]{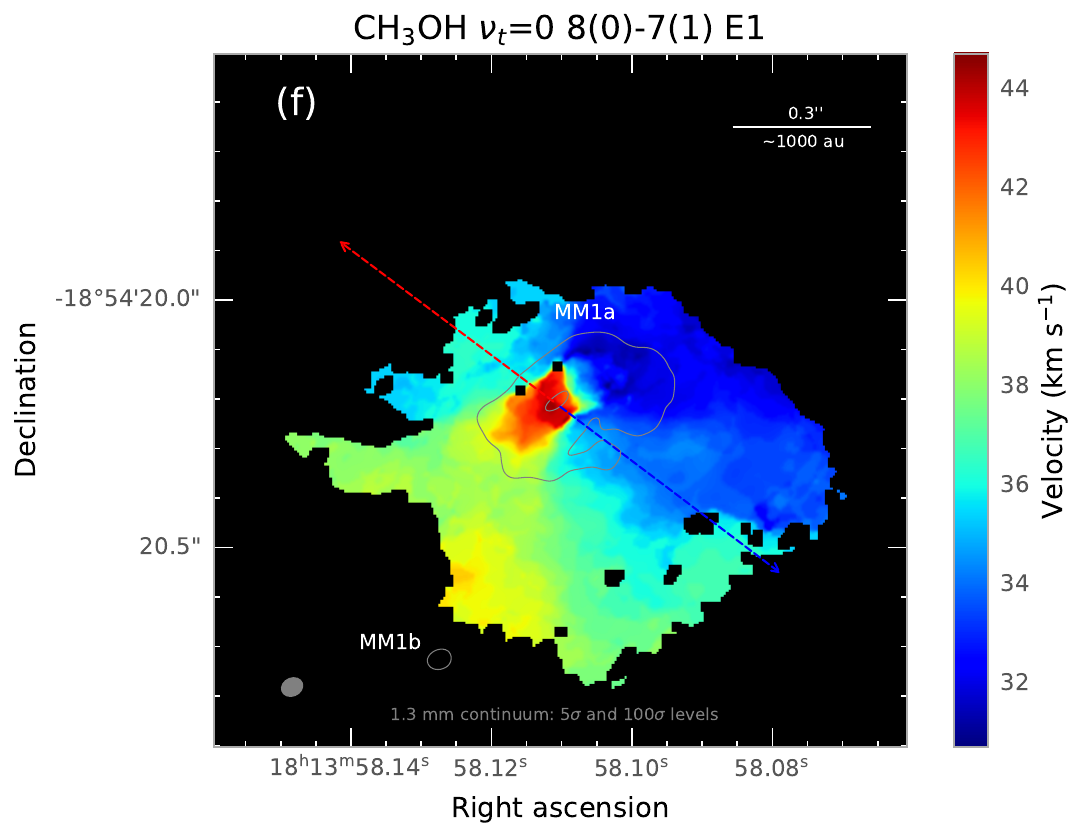}  \\
  \includegraphics[width=60mm]{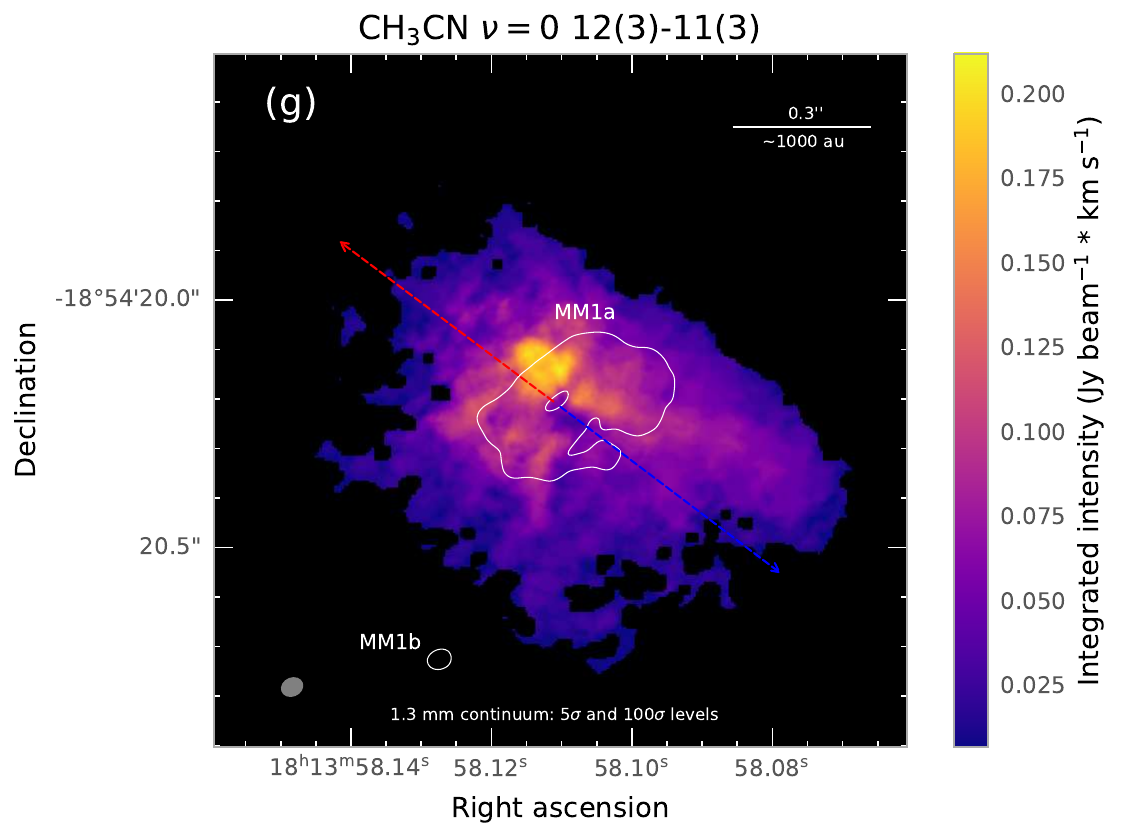}  &
  \includegraphics[width=60mm]{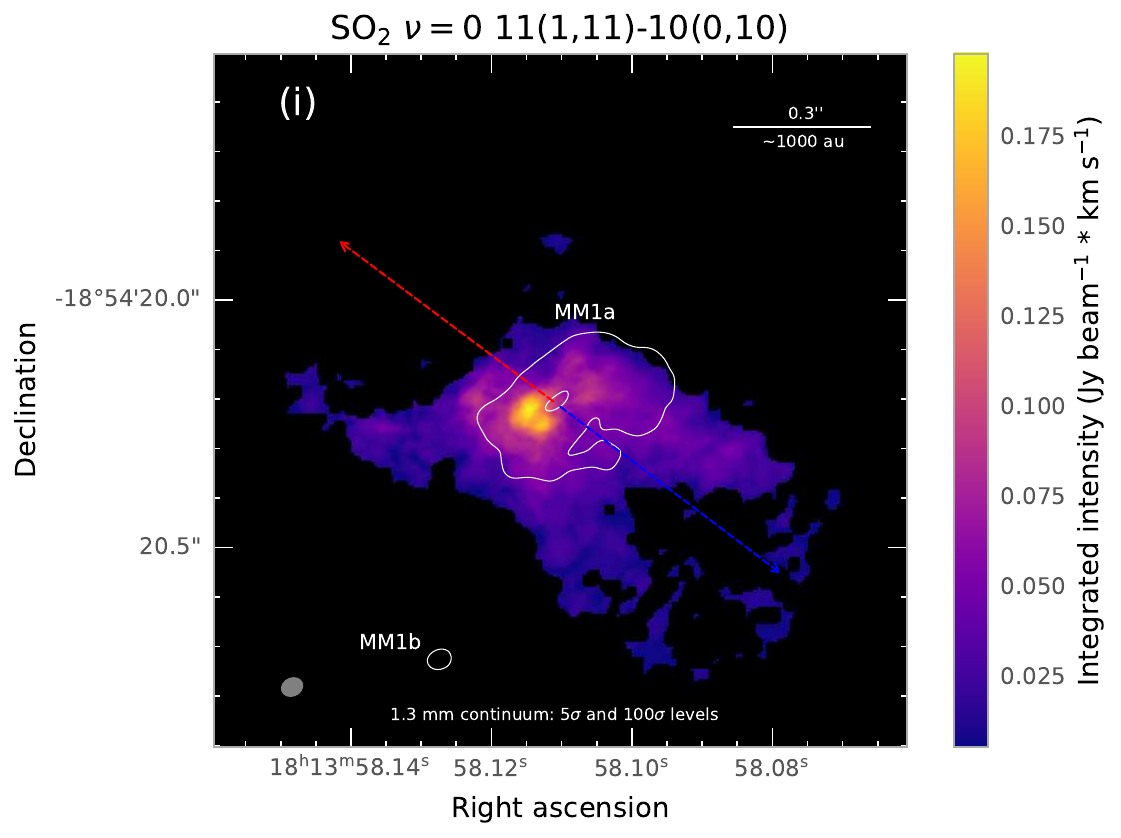}  &
  \includegraphics[width=60mm]{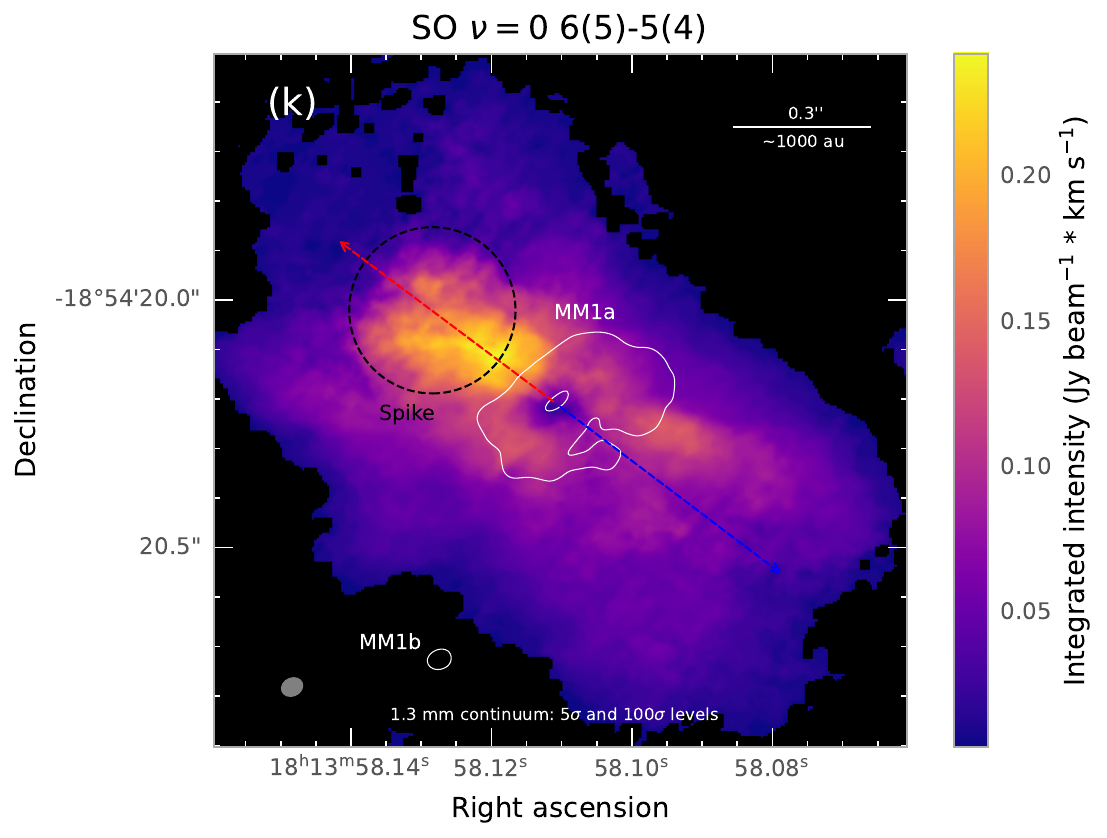}  \\
  \includegraphics[width=60mm]{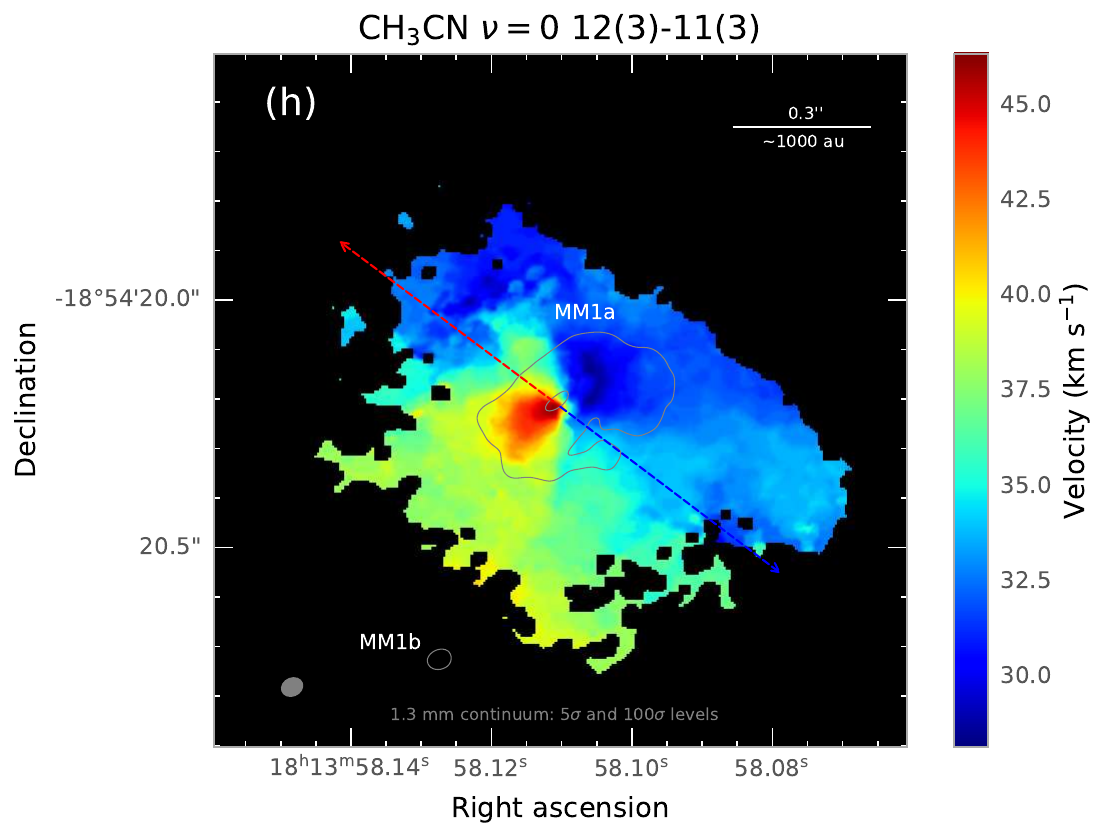}  &
  \includegraphics[width=60mm]{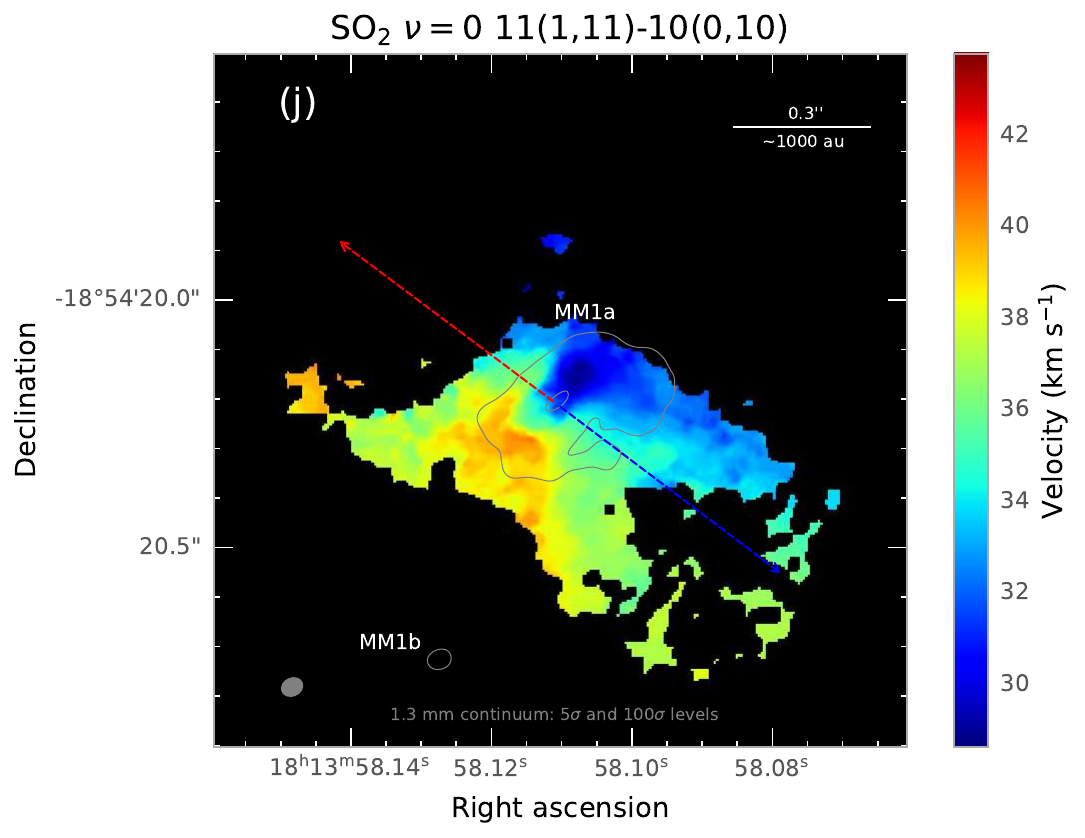}  &
  \includegraphics[width=60mm]{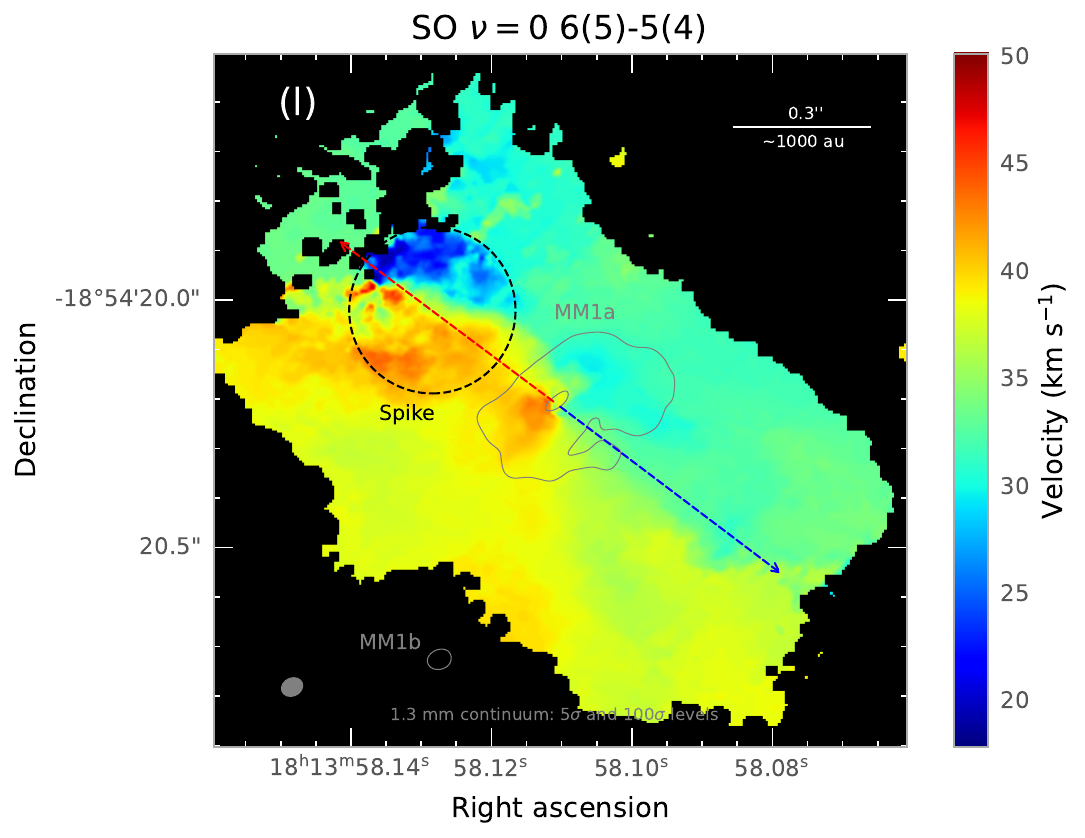}  \\
\end{tabular}
\caption{Moment 0 and 1 maps for the selected molecular line emissions detected at different scales. The synthesised beam of the molecular line data is indicated as a grey ellipse in the lower-left corner. The detected 1.3 mm continuum emission is indicated by white and grey contours (the levels are indicated at the bottom of each panel). The blue and red arrows indicate the  orientation of the large-scale $^{12}$CO outflow presented in \cite{Ilee2016}. The dotted black circle in the SO maps (panels k and l) indicates the velocity spike.
\label{fig:M0M1}}
\end{figure*}

Abundant molecular line emission is detected towards G11.92-0.61 MM1a.
In order to illustrate the molecular emission detected at different physical scales, we present integrated intensity (moment 0) and velocity field (moment 1) maps for selected molecular lines (see Fig. \ref{fig:M0M1}). 

All detected molecular lines are found to be associated with the source MM1a, no emission is detected at the position of the MM1b continuum peak. 
The most compact emissions of the high-excitation energy CH$_3$CN $\varv_8=1$ and $\varv=0$ from $K$=4 to 8  molecular lines are found to originate from a $\sim$0$\farcs$2 (or $\sim$700 au) region around MM1a, showing a similar size, elongation, and PA as the 1.3 mm continuum emission (Figs. \ref{fig:M0M1}a and \ref{fig:M0M1}c). 
Apart from the compact region around MM1a, the lower-excitation energy lines of CH$_3$OH and CH$_3$CN $\varv=0$ $K$=1 to 3 also follow a larger region with an extent of $\sim$0$\farcs$6 ($\sim$2000 au) in the direction perpendicular to the continuum source and the compact molecular line emission (Figs. \ref{fig:M0M1}e and g). 
Finally, the SO$_2$ and SO emissions is elongated up to $\sim$1$\arcsec$ ($\sim$3000 au) along the outflow axis (Figs. \ref{fig:M0M1}i and \ref{fig:M0M1}k).

\begin{table*}
\caption{Parameters obtained from the detected spectral lines.\tablefootmark{a}}
\label{tab:l-param}
\centering  
\begin{tabular}{lcccc}
\hline\hline   
\multicolumn{1}{c}{Lines used\tablefootmark{(b)}}   &  
 N\tablefootmark{(c)} & T$_{ex}$\tablefootmark{(c)} & log(N/N(H$_2$))\tablefootmark{(d)} & log(N$_{\rm H_2})$\\ 
 &  $\times$10$^{15}$ cm$^{-2}$ & K & log cm$^{-2}$ & log cm$^{-2}$ \\
\hline 
CH$_3$CN $\varv_8=1$ $J$ = 12-11 $K$ = -5 to 8 & 5$\pm$0.2 & 333$\pm$21 & -8.7$\pm$1.1 & 24.4$\pm$1.1 \\
CH$_3$CN $\varv=0$ $J$ = 12-11 $K$ = 4 to 9 &  4$\pm$0.1 & 327$\pm$19 & -8.7$\pm$1.1 & 24.3$\pm$1.1 \\
CH$_3$OH $\varv_t=0$ 8$_{0,8}$-7$_{1,6}$ E &  500$\pm$80 & 285$\pm$15 & -7.3$\pm$1.0 & 26.4$\pm$1.0\\
CH$_3$CN $\varv=0$ $J$ = 12-11 $K$ = 0 to 3 &  6$\pm$0.3  & 248$\pm$10 & -8.7$\pm$1.1 & 24.5$\pm$1.1 \\
\hline  
\end{tabular}
\tablefoot{$^{a}$For a position $\sim$0$\farcs$05 to the SW of MM1a, see Sect. \ref{ResLines}.
$^{(b)}$Throughout the paper we use the nomenclature of the JPL catalogue. 
$^{(c)}$Determined using the \textit{SLIM} task of the MADCUBA package by simultaneously fitting multiple transitions of CH$_3$CN $\varv_8=1$ (the whole spectrum), CH$_3$CN $\varv=0$ (low and high-K transition fitted separately), and CH$_3$OH (including the weaker transitions 25$_{3,22}$-24$_{4,20}$ and 23$_{5,19}$-22$_{6,17}$) under the assumption of local thermodynamic equilibrium; the fitting process directly adjusts T$_{ex}$ and N to reproduce the observed line intensities and their ratios, without assuming a prior relation between T$_{ex}$ and the column density.
$^{(d)}$The abundances are calculated from the average column densities reported in \cite{Gieser2021}.} 
\end{table*}

To estimate the physical parameters traced by different molecular lines, we selected an area of the size of the synthesised beam ($\sim$30 mas) shifted by $\sim$0$\farcs$05 to the SW of the continuum peak, as the fit towards the continuum peak  turned out to be complicated by the presence of multiple kinematic components at this position.
Then for each data cube we extracted spectra integrated over the selected area with the Spectral Profile plotting tool of the CASA image viewer. The spectra were fitted using the SLIM tool of MADCUBA.
We fitted all 14 lines of CH$_3$CN $\varv_8=1$ simultaneously, while low ($K$ = 0 to 3) and high-$K$ ($K$ = 4 to 9) CH$_3$CN $\varv=0$ transition were fitted separately to take into account the variation of excitation temperatures with $K$ since they trace different environments (Fig. \ref{fig:M0M1}c vs \ref{fig:M0M1}g). The CH$_3$OH $\varv_t=0$ 8$_{0,8}$-7$_{1,6}$ E transition was fit together with two weaker CH$_3$OH $\varv_t=0$ lines, 25$_{3,22}$-24$_{4,20}$ and 23$_{5,19}$-22$_{6,17}$ (however, as these two lines are weak, we do not use them in imaging). We note that we could fit only those molecular species for which there was more than one  line in the spectrum, a condition necessary to derive excitation temperature and column density estimates (only single lines were detected for SO and SO$_2$).
Physical parameters for the fitted molecular species are presented in Table \ref{tab:l-param}.
To estimate the physical conditions of the gas, we employed the SLIM task of the MADCUBA package and not the rotational diagram method, since the latter assumes optically thin emission and a single excitation temperature. In contrast, SLIM performs a non-linear fit to the observed spectra, accounting for line opacity, beam dilution, and line blending effects, thereby offering a more reliable characterisation of the molecular excitation conditions.

Comparison of the parameters obtained for different molecular lines
(Table \ref{tab:l-param}) shows that compact CH$_3$CN $\varv_8=1$ and  $\varv=0$ $K$ = 4 to 9 emissions trace a slightly hotter gas (T$_{ex}$$\approx$330 K) located deeper in the core.
While, as expected for gas located at greater separation from the central source, CH$_3$OH and CH$_3$CN $\varv=0$ $K$ = 0 to 3 transitions indicate slightly lower temperatures (T$_{ex}$$\approx$250 K). 
The column density estimates derived for the CH$_3$CN and CH$_3$OH molecules are consistent, within the uncertainties, with the mean column densities reported in the CORE survey of high-mass star-forming regions by \cite{Gieser2021}. Thus, the physical and chemical conditions in our target region align well with those observed in a broader sample of high-mass star-forming environments.
To estimate the H$_2$ column densities, we compared our estimated column densities of CH$_3$CN and CH$_3$OH listed in Table \ref{tab:l-param} with the abundances of these molecules calculated based on the typical column densities presented in Fig. 8 of \cite{Gieser2021}. 
The obtained values of $\sim$10$^{24}$ cm$^{-2}$  from CH$_3$CN and $\sim$10$^{26}$ cm$^{-2}$ from CH$_3$OH are higher than that  based on the flux density of the 1.3 mm continuum. 
This finding might be explained by considering the different regions traced by molecular gas with respect to the dust continuum. Concentration of dust closer to the central source with gas being driven to larger radii is a predicted feature of accretion disks in both low-mass (for example, \citealt{Perez2012, Vorobyov2018}) and high-mass (for example, \citealt{Seifried2011}) protostars, and was first noted for the disk in G11.92-0.61 MM1 by \cite{Ilee2016}. Additionally, the larger H$_2$ column density calculated for CH$_3$OH likely reflects the fact that the CH$_3$OH gas occupies a larger region (and thus a larger `column' of gas) than CH$_3$CN. 

All velocity field  maps (Figs. \ref{fig:M0M1}b, \ref{fig:M0M1}d, \ref{fig:M0M1}f, \ref{fig:M0M1}h, \ref{fig:M0M1}j, and \ref{fig:M0M1}l) show a clear velocity gradient within the dust continuum region around MM1a consistent with rotation around the source and supporting the presence of a disk as previously proposed by \cite{Ilee2016, Ilee2018}.
In addition to the disk, all the lines also trace gas in the outflow direction (indicated by red and blue arrows in Fig. \ref{fig:M0M1}), albeit to different extents. 
 `Strings' of emission coming from MM1a to the east, west, and south, seemingly outlining an outflow cavity, are clearly seen in the SO$_2$ $\varv=0$ map  (Fig. \ref{fig:M0M1}k), and can also be distinguished in all other lines.

Remarkably, the velocity field  maps (Figs. \ref{fig:M0M1}b, \ref{fig:M0M1}d, \ref{fig:M0M1}f, \ref{fig:M0M1}h, \ref{fig:M0M1}j, and \ref{fig:M0M1}l) at all the probed scales also suggest the presence of rotation around the direction approximately perpendicular to the major axis of the continuum source (in other words, around the outflow axis; red and blue arrows in Fig. \ref{fig:M0M1}), with redshifted velocities to the south-east and blueshifted velocities to the north-west.
The presence of a clear transversal V$_{\rm LSR}$ gradient at all positions suggests the existence of a large-scale rotation along the outflow.
One more surprising feature of the velocity field  maps is the presence of a velocity spike in the SO Moment 1 map (see the dotted black circle in Fig. \ref{fig:M0M1}l). In contrast to all other imaged molecular lines, in the SO moment 1 map, a region with the reddest and bluest velocities is found not at the position of MM1a but at a separation of $\sim$0$\farcs$3 ($\sim$1000 au) to the NE of it.

\subsection{Position-velocity diagrams}

\begin{figure*}
\centering
\begin{tabular}{ccc}
  \includegraphics[width=60mm]{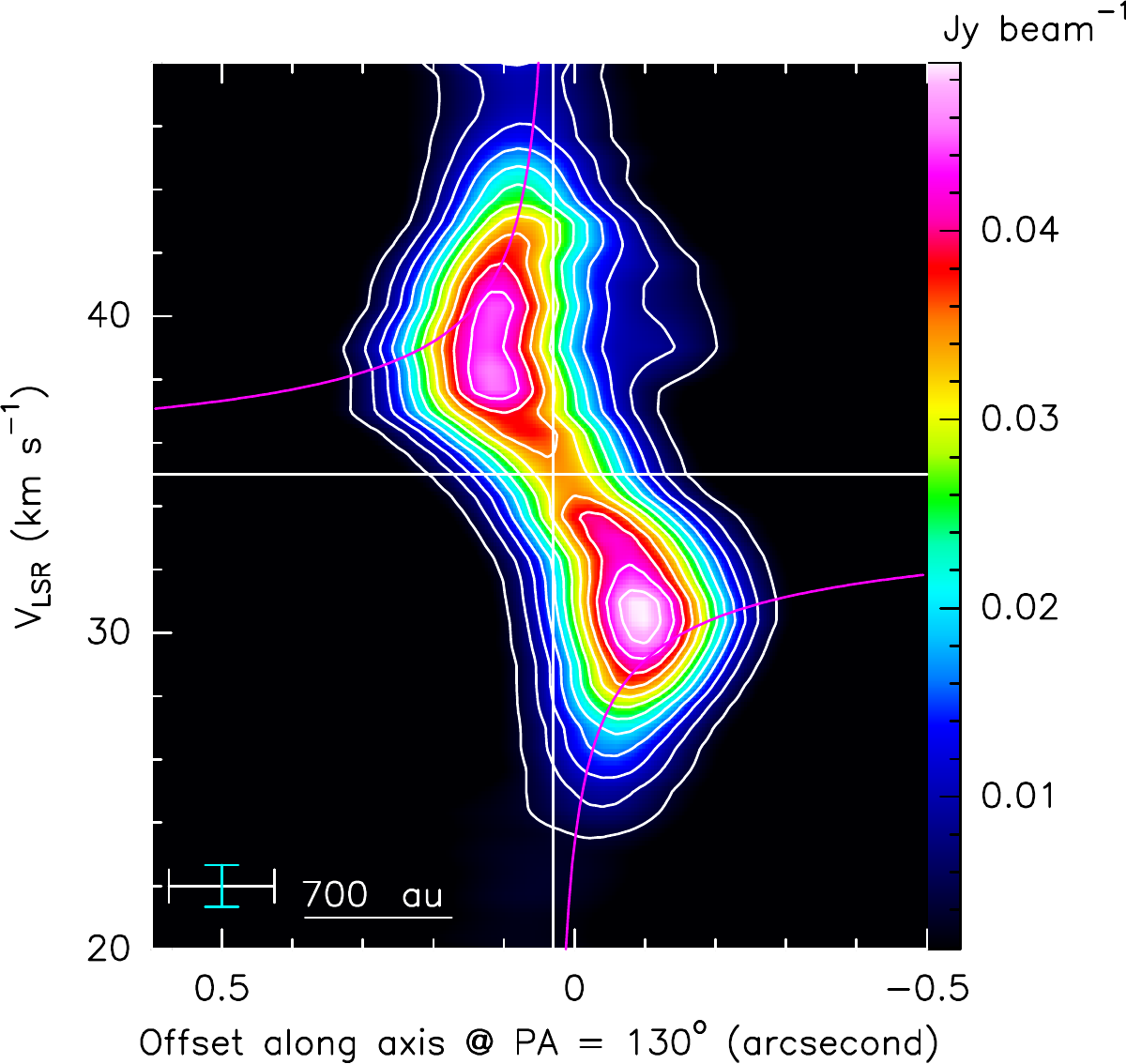}  &
  \includegraphics[width=60mm]{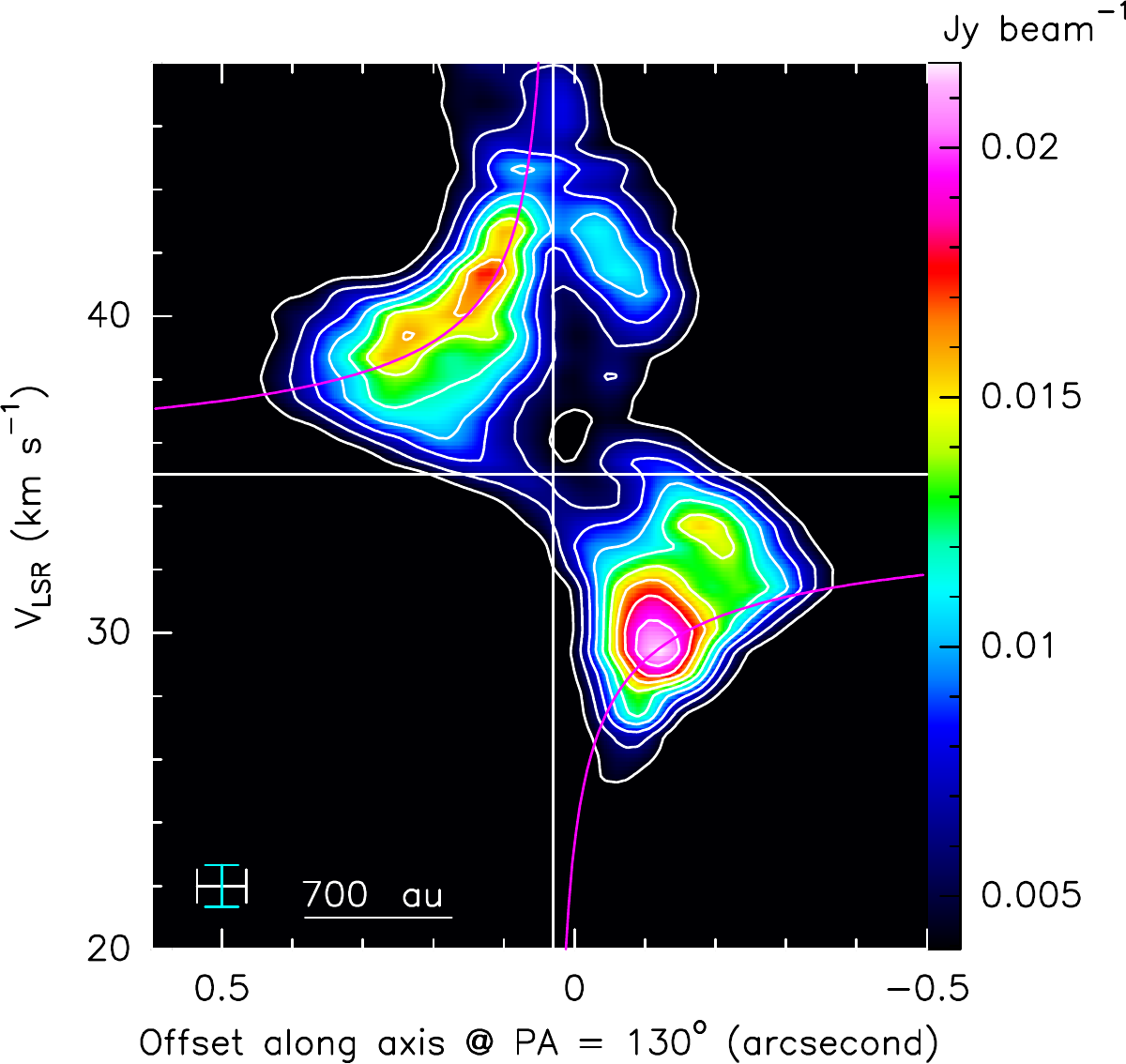}  &
  \includegraphics[width=60mm]{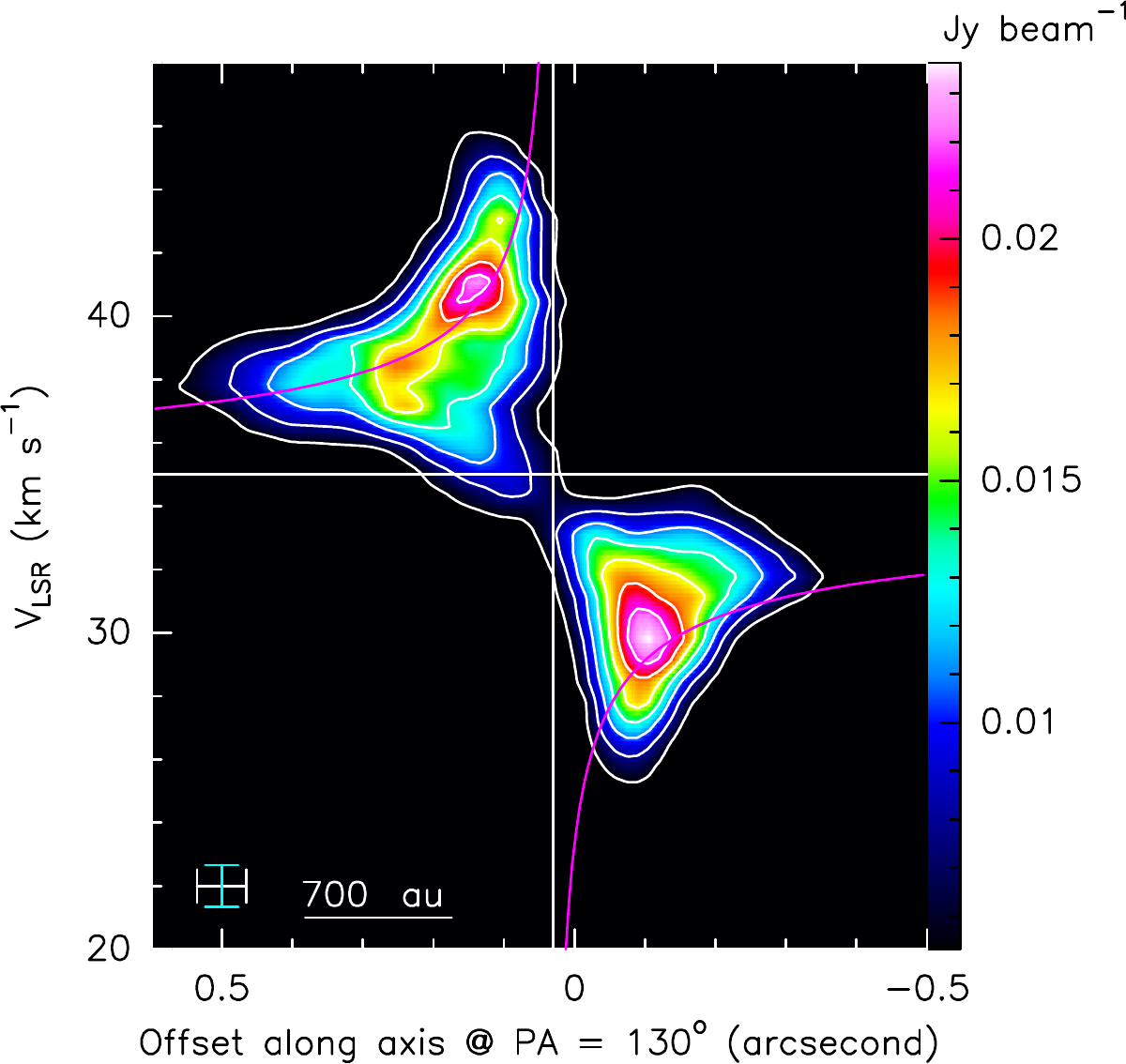}  \\
(a)  & (b) & (c)  \\[6pt]
\end{tabular}
\caption{PV diagrams of the CH$_3$CN $\varv=0$ $J$ = 12-11 $K$ = 7 (a), CH$_3$OH $\varv_t=0$ 8$_{0,8}$-7$_{1,6}$ E (b), and SO (c) emission along the disk. The magenta curves represent the Keplerian disk model with a central mass of 15 $M_{\odot}$.
\label{fig:pv}}
\end{figure*}

We analysed the velocity profile  along the disk plane (along the major axis of the continuum emission with PA $\simeq$ 130$^{\circ}$) using the PV diagrams presented in Fig. \ref{fig:pv}. 
Three lines were selected to be representative of the three identified emission scales: CH$_3$CN $\varv=0$ $K$=7 for the compact scale, CH$_3$OH for the intermediate, and SO for the large-scale emission.

While all three selected tracers show similar symmetrical double-peaked profiles, they probe different extents and kinematics of the disk. The CH$_3$CN profile has the steepest velocity gradient and seems to deviate from the Keplerian profile at larger radii from the central star.
The CH$_3$OH and SO emissions trace larger radii than the CH$_3$CN line, and their profiles appear to be Keplerian-like over the whole radial extent.

\subsection{Keplerian disk approximation} \label{Kep}

\begin{figure}
\centering
  \includegraphics[width=90mm]{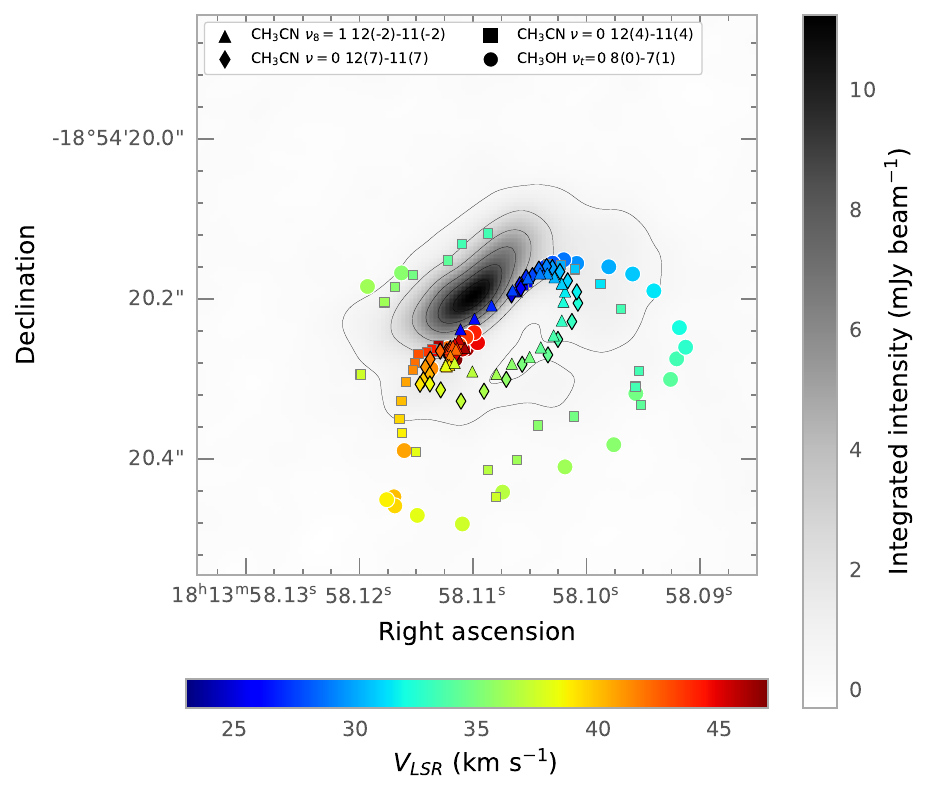}  
\caption{Combined ALMA peak map for compact molecular emission detected in the vicinity of MM1a (see the legend). The grey-scale image and contours represent the 1.3 mm ALMA continuum (same as in Fig. 1). The markers are coloured by velocity (see the colour bar at the bottom).
\label{fig:spot}}
\end{figure}

To further analyse the distribution and kinematics of the compact molecular emission in the vicinity of MM1a (in other words, the emission associated with the rotating disk), we fitted with a 2D Gaussian the emission peaks in each channel of the data cubes of six spectral lines: CH$_3$CN $\varv_8=1$ $J$ = 12$-$11 $K$ = -2 and 2; CH$_3$CN $\varv=0$ $J$ = 12$-$11 $K$ = 4, 7, and 8; and CH$_3$OH $\varv_t=0$ 8$_{0,8}$-7$_{1,6}$ E. 
This method was previously used in, for example, \cite{Moscadelli2019}.
The fitted transitions were selected based on their (1) compactness (the emission should have a defined peak), 
(2) isolation in the spectrum (heavily blended lines were excluded from the analysis), (3) intensity (peak flux density above 5$\sigma$). This approach allows us to locate the molecular emission with a high positional accuracy.
The resulting combined peak map is presented in Fig. \ref{fig:spot} (note that in order to avoid overcrowding of the map, we excluded CH$_3$CN $\varv_8=1$ $J$ = 12$-$11 $K$ = 2 and CH$_3$CN $\varv=0$ $J$ = 12$-$11 $K$ = 8 as they closely follow the distribution of CH$_3$CN $\varv_8=1$ $K$ = -2 and CH$_3$CN $\varv=0$ $K$ = 7, respectively).

Independently of the  tracer, all emission peaks follow the same pattern with blueshifted velocities to the north-west and redshifted velocities to the south-east (Fig. \ref{fig:spot}). The peaks with extreme (bluest and reddest) velocities (V$_{\rm LSR}$=$\sim$25-30 and 39-47 km s$^{-1}$, respectively) are spread along a line parallel and close to the major axis of the continuum source underlining a clear velocity gradient.

Assuming that the obtained peak map represents a rotating disk, we can determine the mass of the central object and the geometry of the disk. To do so, we fitted the peak distribution (Fig. \ref{fig:spot}) 
with two models: (1) a linear Keplerian model that takes into account only high velocity blue- and redshifted peaks (V$_{\rm LSR}$=$\sim$25-30 and 39-47 km s$^{-1}$) that draw a linear structure at the position of MM1, (2) a planar Keplerian model \citep{Sanchez2013} that fits all peaks and provides, in particular, an estimate of the inclination angle of the disk.

\begin{figure}
\centering
\begin{tabular}{cc}
  \includegraphics[width=69mm]{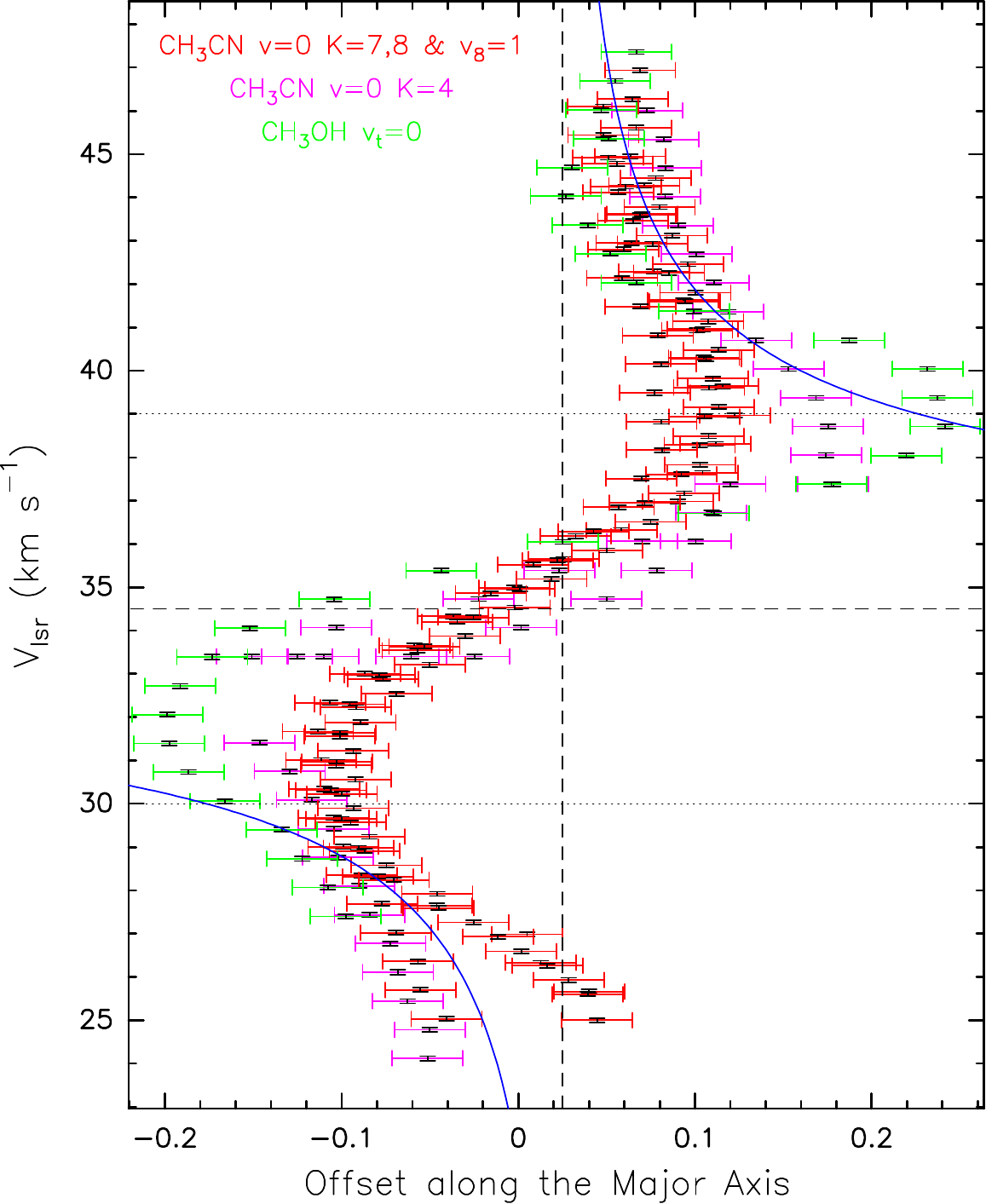}  
\end{tabular}
\caption{Linear Keplerian model. Red, magenta, and green error bars give major-axis-projected positions and V$_{\rm LSR}$ for the high-velocity peaks (the velocity cuts are indicated by horizontal dotted lines) of the CH$_3$CN and CH$_3$OH molecular data (see the legend). Dashed black lines indicate the position of the central source and its V$_{\rm LSR}$. The blue curves represent a Keplerian fit for the central mass of 15 $M_{\odot}$.}
\label{fig:Kep-lin}
\end{figure}

The linear Keplerian fit is illustrated in Fig. \ref{fig:Kep-lin}, which 
shows the V$_{\rm LSR}$ of the emission peaks from Fig. \ref{fig:spot} versus the corresponding offsets projected along the disk major axis (along PA=130$^{\circ}$).
The offset position of the central source along the X-axis ($X_0$), the systemic velocity (V$_{\rm star}$), and central source mass (M) were the free parameters of the linear Keplerian fit, whose result is reported in Table \ref{tab:disk}. 
The linear Keplerian fit provides a position of the central source slightly offset by 0$\farcs$03 from the position of the continuum peak that we used as the reference and V$_{\rm star}$=34.5 km s$^{-1}$. The fitted mass of the central source is  $\sim$15 $M_{\odot}$ without consideration of the disk inclination angle. 

\begin{figure}
\centering
  \includegraphics[width=90mm]{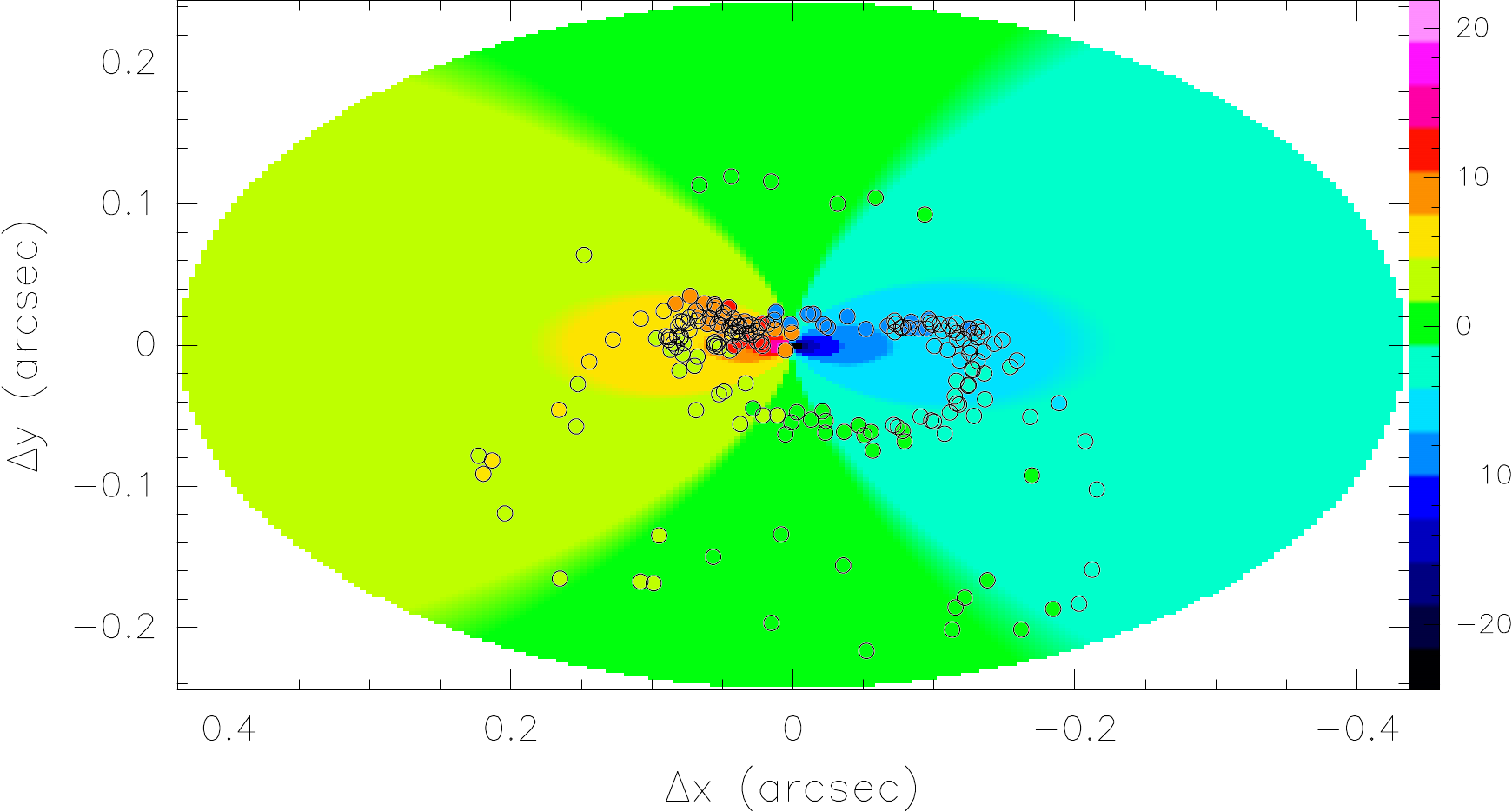}  
\caption{Planar Keplerian fit (considering all velocity features) of the CH$_3$CN and CH$_3$OH molecular data. The modelled velocity field is indicated by colour (the colour match of the spot with the corresponding sector of the velocity field indicates a correct approximation). The map is rotated by 50$\degr$ clockwise with respect to that in Fig. \ref{fig:spot}. 
\label{fig:Kep}}
\end{figure}

The  free parameters of the planar Keplerian fit were:
the position of the central source on the plane of the sky ($X_0$, $Y_0$), the central velocity (V$_{\rm star}$), the PA of the disk projection on the plane of the sky (PA), the inclination ($\theta$) of the disk rotation axis with respect to the plane of the sky, and the central source mass (M). 
Interestingly, if we use only the emission peaks that make up the inner ellipse around MM1a (all lines excluding the more extended emission of CH$_3$CN $\varv=0$ $J$ = 12-11 $K$ = 4 and CH$_3$OH), we obtain an inconsistent fit with the linear distribution of high-velocity spots offset from the disk's major axis and a large disk inclination of $\sim$56$^{\circ}$.
Inclusion of the more extended CH$_3$CN $\varv=0$ $K$ = 4 and CH$_3$OH data, yield the fit in Fig. \ref{fig:Kep} where the modelled velocity field (sectors of different colour) corresponds well to the data. 
The best-fit parameters are given in Table $\ref{tab:disk}$.
We have ascertained that the inconsistent planar Keplerian fit obtained by employing only the CH$_3$CN data (excluding $\varv=0$ $K$ = 4)  is due to the mid-velocity spots (V$_{\rm LSR}$ = 30-39 km s$^{-1}$) spreading out of the disk mid-plane (see Fig. \ref{fig:spot}). 
Their kinematics is inconsistent with Keplerian rotation around the same central mass as traced by the linear distribution of the corresponding high-velocity (V$_{\rm LSR}$ $<$30 km s$^{-1}$ and $>$39 km s$^{-1}$) spots.
From the Keplerian planar fit, we obtained the best-fit mass of the central source of $\sim$20 $M_{\odot}$, which is in good agreement with the result of the linear Keplerian fit corrected for a best-fit disk inclination of $\sim$30$^{\circ}$.
Thus, the two approaches provide similar results, confirming the presence of a slightly inclined, rotating disk around MM1a. 

\begin{table}[h]
\caption{Fitted parameters of the disk.}             
\label{tab:disk}      
\centering                          
\begin{tabular}{cccccc}        
\hline\hline                 
X$_0$ & Y$_0$ & V$_{\rm star}$  &  $\theta$\tablefootmark{(a)} & PA & M   \\  
($\arcsec$) & ($\arcsec$) & (km s$^{-1}$) & ($^{\circ}$)  &  ($^{\circ}$) & ($M_{\odot}$)    \\
\hline                       
\multicolumn{6}{c}{Linear Keplerian fit} \\
\hline 
0.03\tablefootmark{(b)} & - & 34.5  & - & -\tablefootmark{(c)} & 15\tablefootmark{(d)}  \\
\hline                       
\multicolumn{6}{c}{Planar Keplerian fit} \\
\hline                                   
0.01 & -0.03 & 34.5  & 32  & 130 & 20 \\
\hline                       
\end{tabular}
\tablefoot{$^{(a)}$Inclination, i.e. the angle between the disk axis and the plane of the sky. $^{(b)}$Offset along the PV cut. $^{(c)}$Was set to  130$^{\circ}$ in accordance with the PA of the continuum source. $^{(d)}$Not corrected for the disk inclination.} \\
\end{table}

\section{Discussion} \label{discussion}
\subsection{Disk and central mass}

Previous observations of G11.92-0.61 MM1a had revealed the presence of a massive protostar surrounded by an accretion disk; however, the exact parameters of this system remained a subject of debate. Our new high-resolution, multi-line ALMA data provided the insights needed to resolve these uncertainties and offer a more precise estimate of the characteristics of the central object and its disk.

The work of \cite{Ilee2016} suggested the presence of an infalling Keplerian disk with a radius of 1200 au and an inclination of $\sim$55-38$^{\circ}$\footnote{Note that \cite{Ilee2016, Ilee2018} and this work use different definitions of inclination angle: in \cite{Ilee2016, Ilee2018}, 0$^{\circ}$ corresponds to a face on disk while, in this work, 90$^{\circ}$ corresponds to a face-on disk. Thus, we recalculated the inclinations obtained in \cite{Ilee2016, Ilee2018} to express them according to our
definition}. rotating around an enclosed mass of $\sim$30-60 $M_{\odot}$ based on multi-line data obtained with the SMA and \textit{Karl G. Jansky} Very Large Array (VLA). 
Later ALMA observations of CH$_3$CN emission at a higher angular resolution indicated slightly different parameters of $\sim$40 $M_{\odot}$ for the central mass, and a disk inclination of $\sim$20$^{\circ}$ \citep{Ilee2018}.
However, these discrepancies appear to be mostly dependent on the choice of tracers rather than on the resolution of the observations. 
Similar to our analysis, \cite{Ilee2016} also noticed that the Keplerian approximation depends on whether methanol is included or not in the input dataset. In \cite{Ilee2016}, inclusion of the CH$_3$OH data provided a central mass of $\sim$35 $M_{\odot}$ and a disk inclination of $\sim$38$^{\circ}$, while exclusion of the CH$_3$OH yielded $\sim$60 $M_{\odot}$ and $\sim$55$^{\circ}$. The Keplerian fit of our data provided a central mass of $\sim$20 $M_{\odot}$ and disk inclination $\sim$32$^{\circ}$ with CH$_3$OH and $\sim$35 $M_{\odot}$ and disk inclination $\sim$55$^{\circ}$ without it (see Sect. \ref{Kep}). 
Overall, these estimates suggest two scenarios: (1) a more massive protostar with a more inclined disk, or (2) a less massive protostar with a less inclined disk.

While \cite{Ilee2016} could not discriminate between the two possibilities
and confined the mass in the range $\sim$35-60 $M_{\odot}$, we argue that inclusion of the CH$_3$OH data provides a more accurate approximation of the disk geometry, and in particular the disk inclination as CH$_3$OH traces large radii of the disk (Fig. \ref{fig:spot}).
The high-excitation energy CH$_3$CN emissions ($\varv_8=1$ and $\varv=0$ $K$=7 to 8) at mid-velocities (30-39 km s$^{-1}$) deviate from Keplerian rotation at disk radii of $>$0$\farcs$1 ($>$300 au) according to our Keplerian fits as discussed in Sect. 3.4.
Figures \ref{fig:spot} and \ref{fig:Kep} show that the mid-velocity emissions of high-excitation CH$_3$CN and the CH$_3$OH lines draw two  `rings' around the central protostar, and that of CH$_3$CN is significantly smaller (de-projected diameter of $\sim$400-500 au) than that of CH$_3$OH (1200-1500 au). According to Fig. \ref{fig:Kep-lin}, the mid-velocity (30-39 km s$^{-1}$) spots of CH$_3$CN, (tracing the smaller ring in Fig. \ref{fig:Kep}) have line-of-sight velocities closer to V$_{\rm star}$ than expected from the Keplerian profile at the corresponding radius; the high-excitation CH$_3$CN emission is thus tracing sub-Keplerian rotation at these positions.
Deviations from Keplerian rotation within the disk (at the radii of $>$300 au in our case) can be due, for example, to magnetic fields (for example, \citealt{Seifried2011}), while the radius at which Keplerian rotation dominates may vary depending on the magnetic field strength and age of the system (for example, \citealt{Seifried2011, Kuiper2011}). 

For instance, for the protostar G23.01-00.41 with a mass comparable to MM1a, ALMA observations of \cite{Sanna2019} showed that sub-Keplerian rotation dominates the accretion disk at radii between 500 and 2000 au with strong infall signatures, whereas centrifugal equilibrium might be reached in the innermost regions only. Notably, because of the higher angular resolution with respect to \cite{Sanna2019}, here we are also able to probe the rotation curve where gas approaches centrifugal equilibrium near the protostar.
We can also glean interesting insights by comparing G11.92-0.61 MM1a to an object at an earlier stage of evolution. The molecular line study by \cite{Miyawaki2022} of the hot molecular core W49N MCN-a provided a compelling example of the transition from a massive, gravitationally unstable envelope to a Keplerian-like disk. Similar to our findings for G11.92-0.61 MM1a, \cite{Miyawaki2022} demonstrate that the choice of molecular tracers significantly impacts the interpretation of rotation curves and the inferred disk parameters. In both sources, distinct radial regions with differing velocity gradients were identified using different molecular lines.
However, in contrast to G11.92-0.61 MM1a, W49N MCN-a remains dominated by a massive, gravitationally unstable envelope. This distinction likely reflects the differing evolutionary stages of the two systems.




Summarising the results obtained for the G11.92-0.61 MM1a system, we confirm the presence of a dust disk traced by the 1.3 mm continuum emission and having a diameter $\sim$0$\farcs$16 ($\sim$500 au) and PA $\sim$130$^{\circ}$ (Fig. \ref{fig:cont}). Molecular emission of high-excitation CH$_3$CN $\varv_8=1$ and $\varv=0$ $K$=7 to 8, trace two distinct regions of the disk: (1) high-velocity emissions trace the inner $<$300 au radii of the disk, which are dominated by Keplerian rotation; and (2) mid-velocity emission at larger disk radii from $>$300 au traces rotation in a sub-Keplerian regime. Overall, the disk is found to be close to edge-on with inclination of $\sim$30$^{\circ}$. The mass of the central object is determined to be $\sim$20 $M_{\odot}$, which agrees with the luminosity of the source of $\sim$10$^4$ L$_{\rm sun}$ \citep{Cyganowski2011, Moscadelli2016, Ilee2018} better than the higher mass estimates obtained in  previous studies.

\subsection{Kinematic profiles of the molecular outflow}

\begin{figure*}[!h]
\centering
\begin{tabular}{cc}
  \includegraphics[width=85mm]{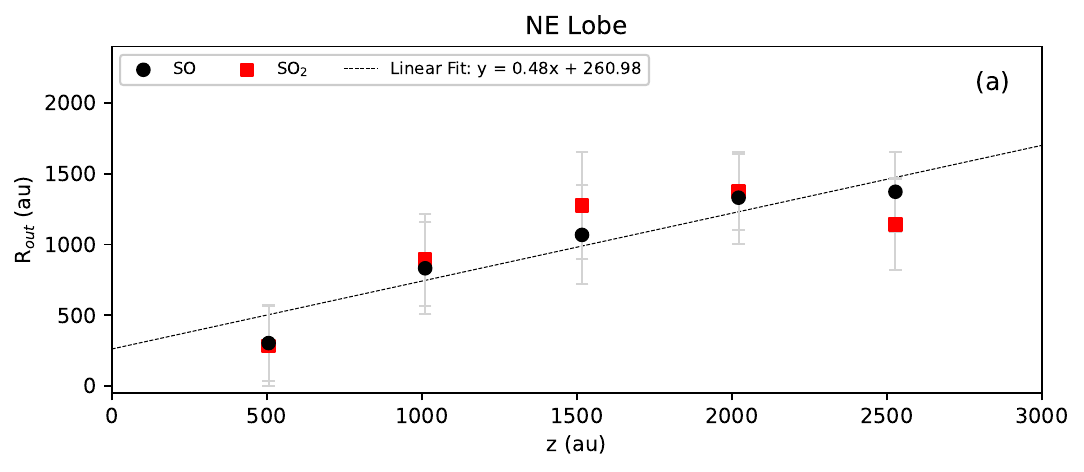}  &
  \includegraphics[width=85mm]{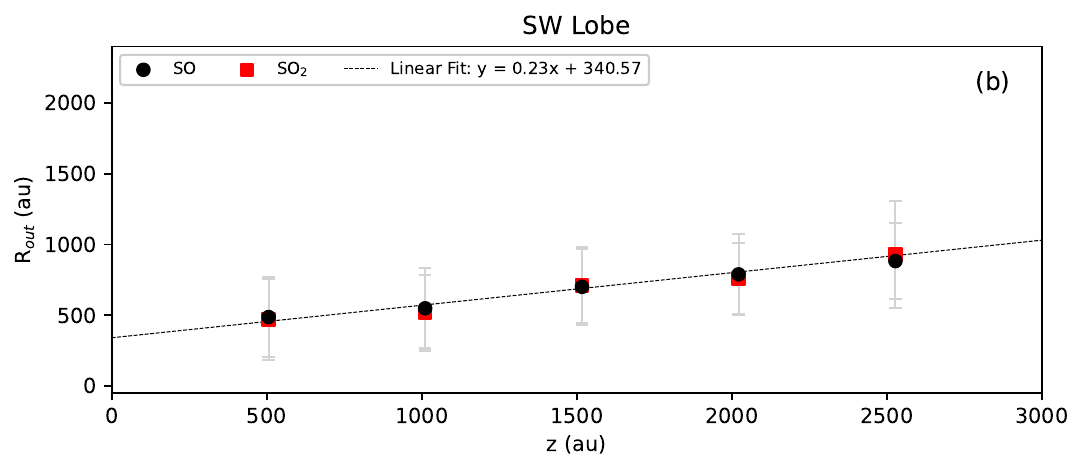}  \\
  \includegraphics[width=85mm]{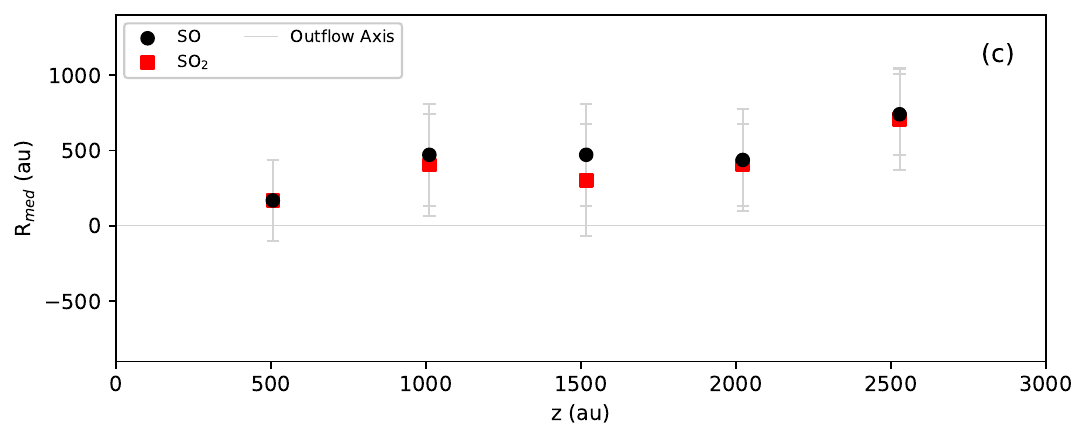}  &
  \includegraphics[width=85mm]{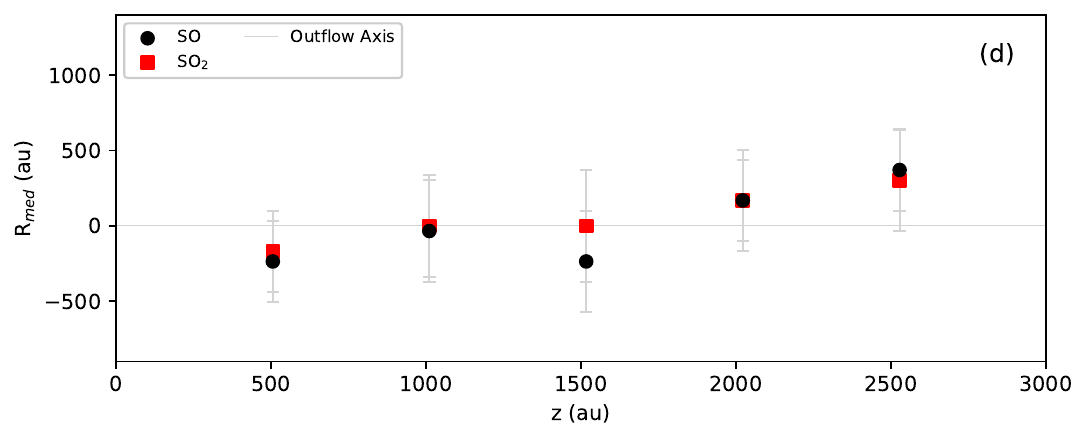}  \\
  \includegraphics[width=85mm]{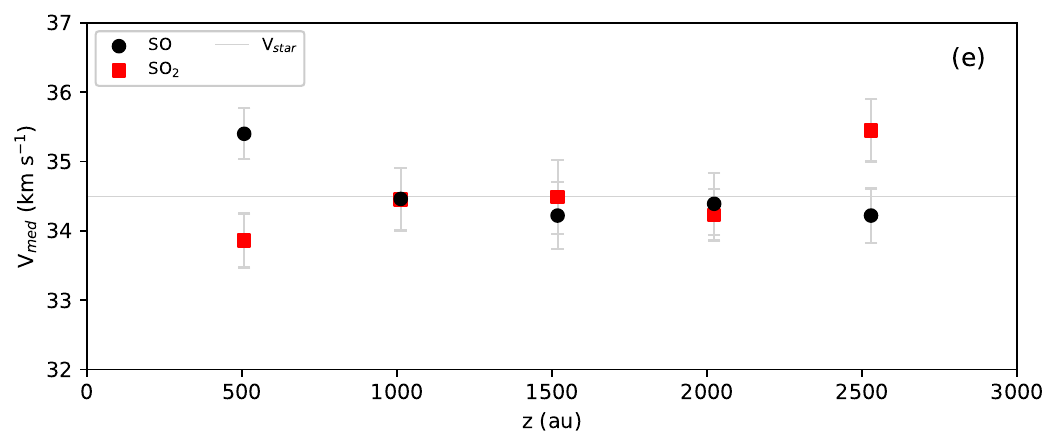} &
  \includegraphics[width=85mm]{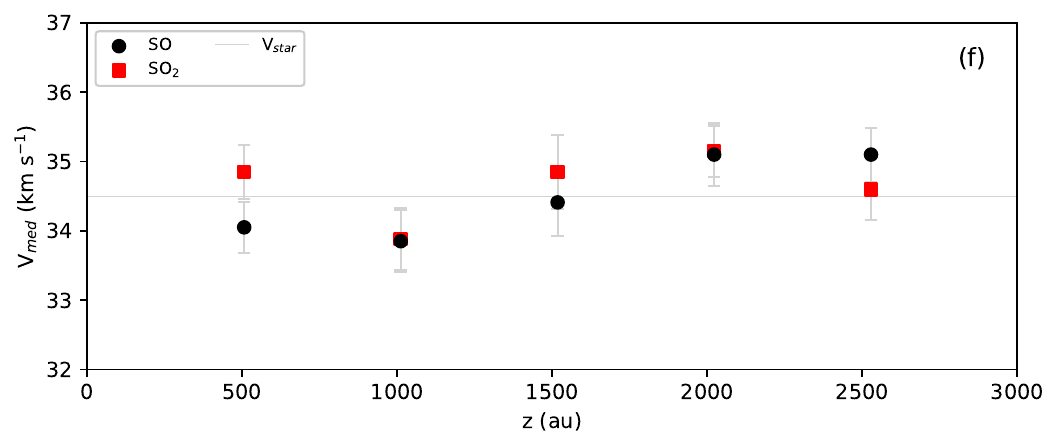} \\
  \includegraphics[width=85mm]{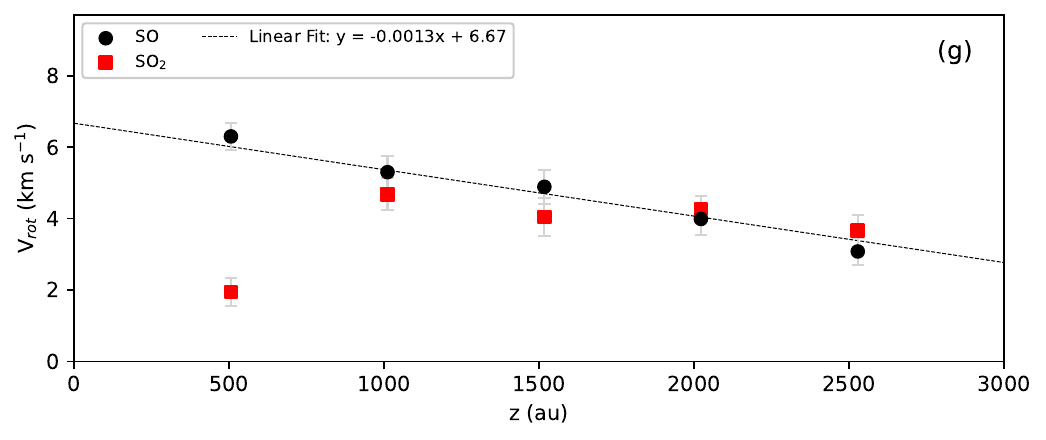} &
  \includegraphics[width=85mm]{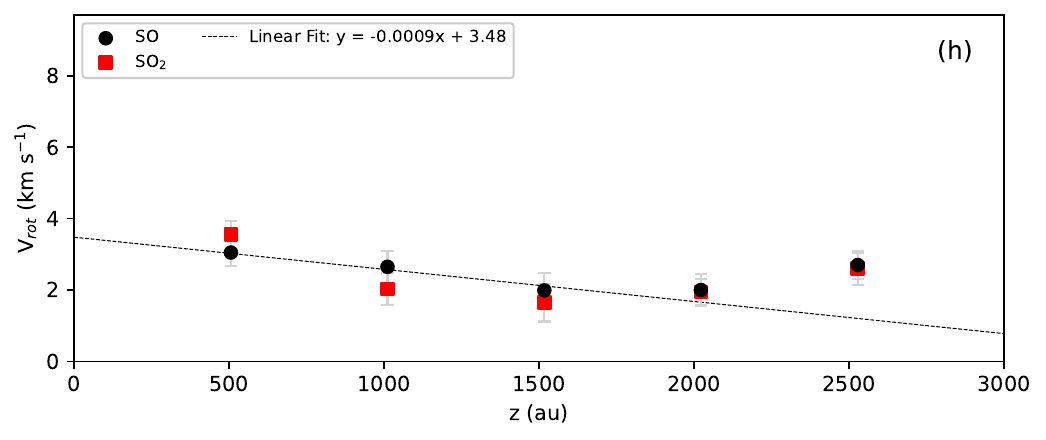}  \\
  \includegraphics[width=85mm]{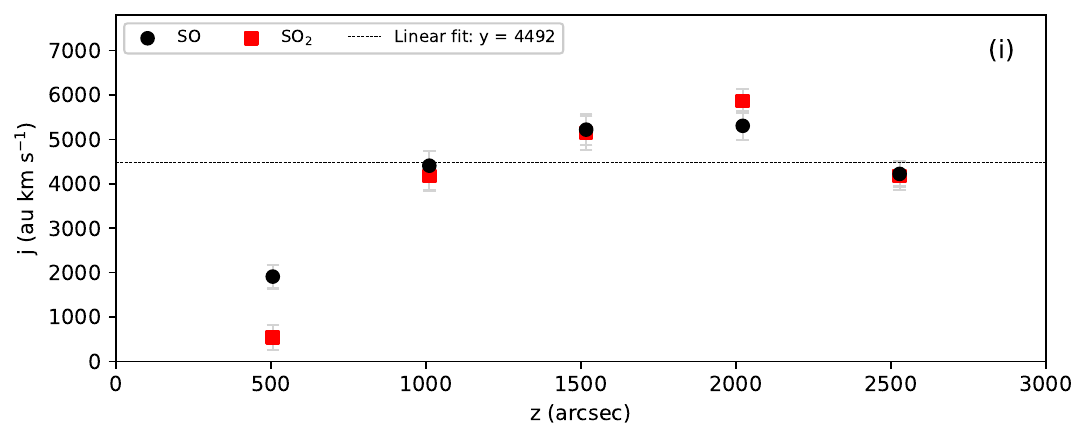}   &
  \includegraphics[width=85mm]{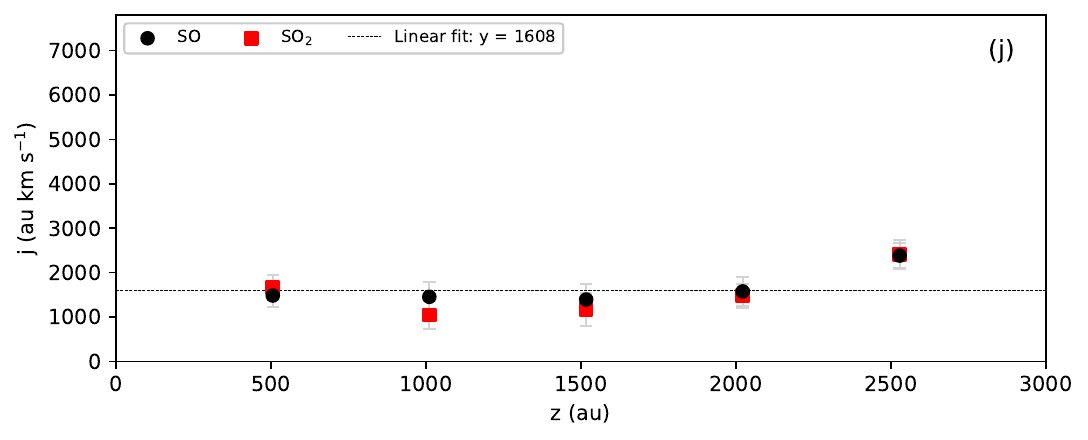}\\
\end{tabular}
\caption{Outflow parameters derived from the PV diagrams for the cuts with steps of 0$\farcs$15 ($\sim$500 au) through the NE (left) and SW (right) lobes of the SO- and SO$_2$-traced outflow. The  markers represent the values obtained using the detected emission peaks in the PV diagrams (see Figs. \ref{fig:pvsone}-\ref{fig:pvso2sw}). Linear fits to the data are indicated with dashed lines (see the legend). }
\label{fig:outf-param}
\end{figure*}

Another insight into the structure of the source is provided by  the low-flux-density 1.3 mm continuum emission, which shows excess emission to the SW of the continuum peak (Fig. \ref{fig:cont}), with two distinct spurs to the eastern and western sides. Notably these spurs of emission pointing to the SW from the continuum peak in our image are roughly symmetrical to the NE excess in the lower resolution image by \cite{Ilee2018} at the same wavelength.
Now comparing the 1.3 mm continuum images from this work and from \cite{Ilee2018} to the molecular line data in Figs. \ref{fig:M0M1}i and \ref{fig:M0M1}k, we see that the SW 1.3 mm emission correlates well with the strings of SO$_2$ and SO emission propagating to the NE and SW from the central source  and tracing outflow motions. Thus, we can infer that the 1.3 mm continuum emission traces not only the disk but also the base of the outflow cavities. And more precisely it traces the dust in the disk and inner flow cavities, as the analysis of the spectral energy distribution from \cite{Ilee2016} demonstrated that the 1.3 mm continuum emission is dominated by thermal dust emission.

All available observations of G11.92-0.61 MM1a indicate the presence of an X-shaped structure parallel to the large-scale outflow and perpendicular to the disk. Besides dust continuum, the structure is clearly seen in our SO$_2$ image (Figs. \ref{fig:M0M1}i-j). It  identified  in all other detected lines (Fig. \ref{fig:M0M1}), as well as in the molecular line data from \cite[especially their HC$_3$N and SO images]{Ilee2016}.
The X-structure could be associated with accretion of material rather than with ejection, that is to say, it could be streamers driving material from the envelope towards the disk, similar to, for example, the case discussed in \cite{Goddi2020}
for the high-mass star-forming complex W51.
However, we discard this possibility as the detected structure is clearly aligned with the large-scale outflow and is clearly seen in outflow tracers (SO$_2$ and SO lines). 
Additionally, the velocity field maps in Fig. \ref{fig:M0M1} hint at an increase in the velocity along the outflow with increased separation from the central protostar, while in the case of streamers, we would expect  higher-velocities as the matter approaches the disk.
Thus, we tend to interpret this structure as  outflowing gas along the outflow cavity.
Moreover, the clear velocity gradient seen in all molecular tracers suggests the presence of rotation of the outflowing gas around the outflow axis.

Assuming that both SO$_2$ and SO emissions trace the walls of the rotating outflow cavities for each of the two outflow lobes, we can try to quantify the outflow properties by examining the kinematic profiles of the molecular emission at different separations from the central source, following 
the methods that have proven to be effective for low-mass protostars (for example, \citealt{Zhang2018}).
Using the SO$_2$ and SO data, we prepared sets of the PV diagrams for several cuts  perpendicular to the outflow axis (z-axis) along the NE and SW lobes (see Figs. \ref{fig:pvsone}-\ref{fig:pvso2sw}). 
Following our model, we presume that in each PV diagram, the peaks of emission in the `red' (redshifted velocities and positive offsets) and `blue' (blueshifted velocities and negative offsets) part of a diagram
correspond to opposite cavity walls of a particular outflow lobe. 
For these two peaks, we measure the offsets from the central source position (R$_{\rm red}$ and R$_{\rm blue}$) and corresponding local standard of rest velocity (V$_{\rm red}$ and V$_{\rm blue}$).
Usage of the lower-resolution data (0$\farcs$15 synthesised beam) allowed us to better determine the peaks of the PV diagrams. 
In accordance with the chosen spatial resolution, the step for the cuts was set to 0$\farcs$15 ($\sim$500 au). For each cut, the radius of the outflow was calculated as R$_{\rm out}$ = (R$_{\rm red}$ $-$ R$_{\rm blue}$)/2 (Fig. \ref{fig:outf-param}a,b), while R$_{\rm med}$ = (R$_{\rm red}$ + R$_{\rm blue}$)/2 is an estimate of  the distance between the cavity walls and the outflow axis (Fig. \ref{fig:outf-param}c,d). The median velocity was estimated as the average  of the velocities V$_{\rm med}$ = (V$_{\rm red}$ + V$_{\rm blue}$)/2 (Fig. \ref{fig:outf-param}e,f) and the rotation velocity was computed as V$_{\rm rot}$ = (V$_{\rm red}$ $-$ V$_{\rm blue}$)/2 (Fig. \ref{fig:outf-param}g,h). We also calculated the specific angular momentum j as R$_{\rm out}$V$_{\rm rot}$ (Fig. \ref{fig:outf-param}i,j). Our errors on the offset and  velocity were determined by the spatial (0$\farcs$15 or 500 au) and spectral (0.35 km s$^{-1}$) resolution, respectively.

Figure \ref{fig:outf-param} shows the parameters of the outflow that we obtained for the NE (left panels) and SW (right panels) lobes using SO (black circle markers) and SO$_2$ (red square markers) data. 
The NE lobe turned out to be 1.5 times wider than the SW lobe, with the maximum R$_{\rm out}$ for the NE lobe being $\sim$1400 au, while the emission from the SW lobe has a maximum R$_{\rm out}$ of $\sim$930 au (Fig. \ref{fig:outf-param}a,b). 
A linear fit to the  R$_{\rm out}$ data provides us with an estimate of the outflow launching radius (R$_0$), which is $\sim$260 au for the NE lobe, and $\sim$340 au for the SW lobe.
The relatively small values of R$_{\rm med}$ with respect to R$_{\rm out}$ indicate that the cavity wall is approximately centred on the outflow axis (Fig. \ref{fig:outf-param}c,d).
SO and SO$_2$ data for  both lobes revealed the same median value of the  velocity V$_{\rm med}$  = 34.4$\pm$0.5 km s$^{-1}$ (grey line in Fig. \ref{fig:outf-param}e and \ref{fig:outf-param}f),
which agrees well with the V$_{\rm star}$ estimated from the Keplerian fits (see Table \ref{tab:disk}). 
Such a good agreement  also suggests that the inclination of the outflow is small.
The rotation velocity (V$_{\rm rot}$) shows a  decreasing trend
with separation from the central source for both the NE and SW lobes (see the dashed line in Fig. \ref{fig:outf-param}g,h). We note that we excluded from the fits the values of V$_{\rm rot}$ (1) obtained for the cut at $\sim$500 au along the NE lobe as the difference between SO and SO$_2$ data points is significant ($\sim$4 km s$^{-1}$) and it seems to indicate that this region closest to the central source is too turbulent or shielded by the dust, and (2) obtained for the cut at $\sim$2500 au along the SW lobe as it seems to reflect a different kinematics in the tail of the outflow.
For the  NE lobe, V$_{\rm rot}$ changes from $\sim$5 km s$^{-1}$ at z =$\sim$1000 au to $\sim$3.4 km s$^{-1}$ at z = $\sim$2500 au; thus, the difference is 1.6 km s$^{-1}$ over 1500 au. 
In the SW lobe, V$_{\rm rot}$ ranges from 3.3 km s$^{-1}$ at z = $\sim$500 au to 2 km s$^{-1}$ at z = $\sim$2000 au, with a variation of 1.3 km s$^{-1}$ over 1500 au, which is very similar to the numbers obtained for the NE lobe.   
Considering the increasing R$_{\rm out}$ of the outflow (Figs. \ref{fig:outf-param}a and \ref{fig:outf-param}b) and decreasing V$_{\rm rot}$ for the cuts that produced reliable estimates, we obtain a distribution of the specific angular momentum (see Figs. \ref{fig:outf-param}i and \ref{fig:outf-param}j) that hints at conservation of this quantity along the flow. In particular,
in the NE lobe, the mean of the specific angular momentum is $\sim$4500$\pm$690 au km s$^{-1}$. For the SW lobe, the mean specific angular momentum is $\sim$1600$\pm$200 au km s$^{-1}$.

The observed conservation of specific angular momentum indicates the presence of differential rotation along the outflow. Such a behaviour is consistent with the expected dynamics of a MHD disk-wind,  where streams of gas, emerge from the disk and co-rotate with their launching points along the poloidal magnetic field lines.
Since SO and SO$_2$ emissions are known to trace denser regions of the outflow, we assume that we detect the rotation of the denser, inner portions of the outflow linked to the disk-wind, rather than the broader, more diffuse outflow cavity.

\subsection{Disk-wind evidence}

\begin{figure*}[!h]
\centering
\begin{tabular}{cccc}
  \includegraphics[width=43mm]{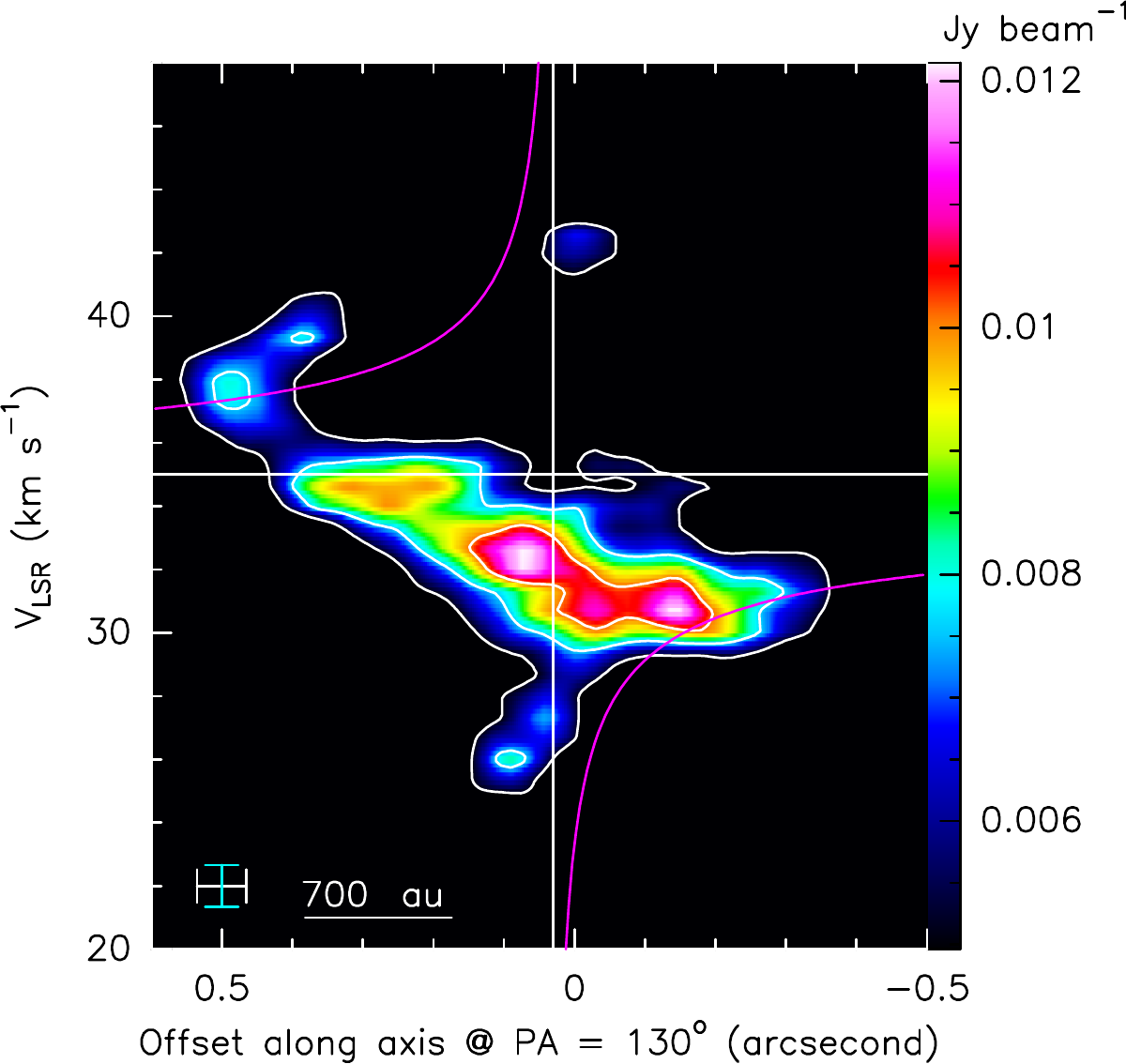}  &
  \includegraphics[width=43mm]{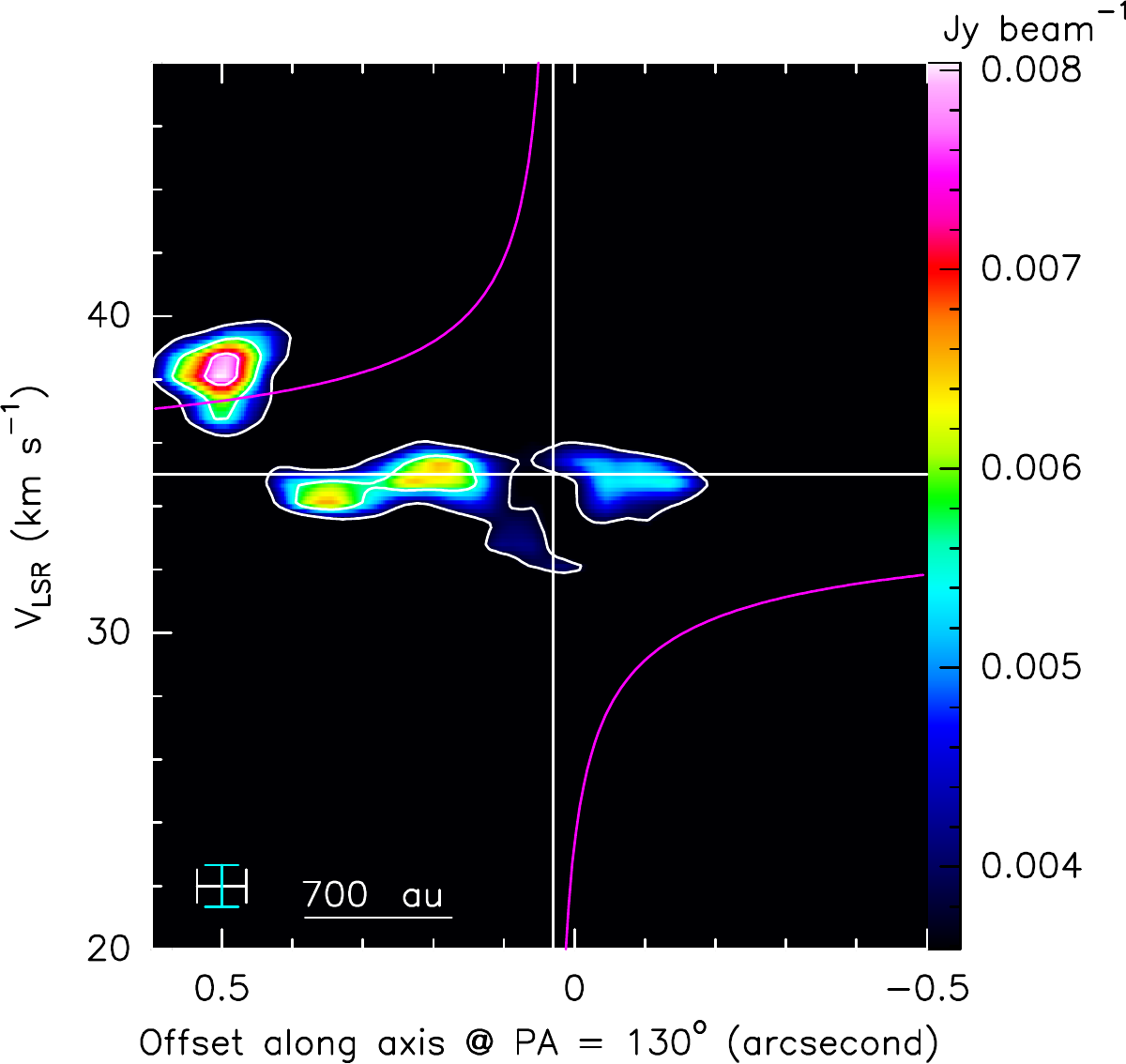}   &
  \includegraphics[width=43mm]{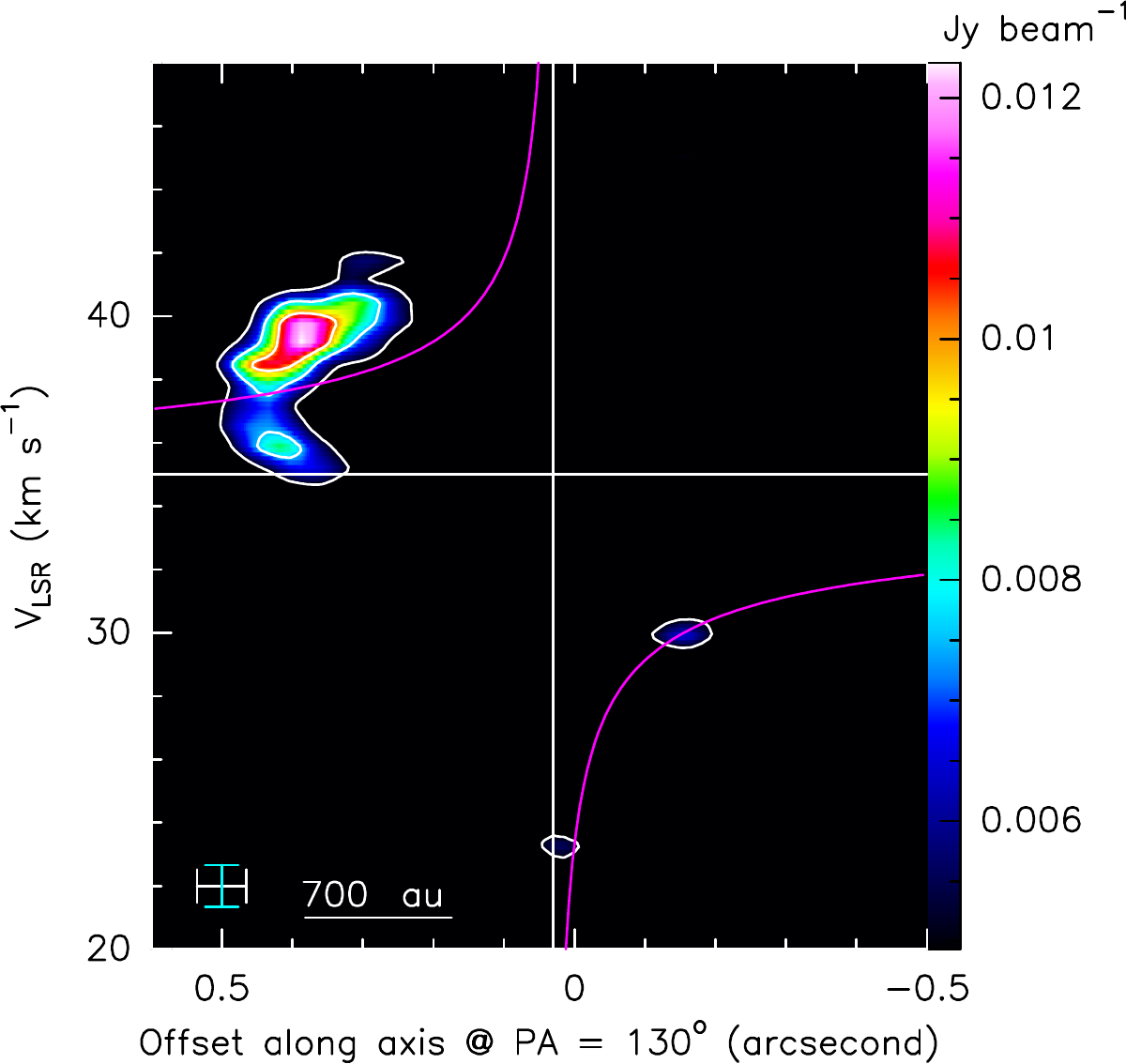}  &
  \includegraphics[width=43mm]{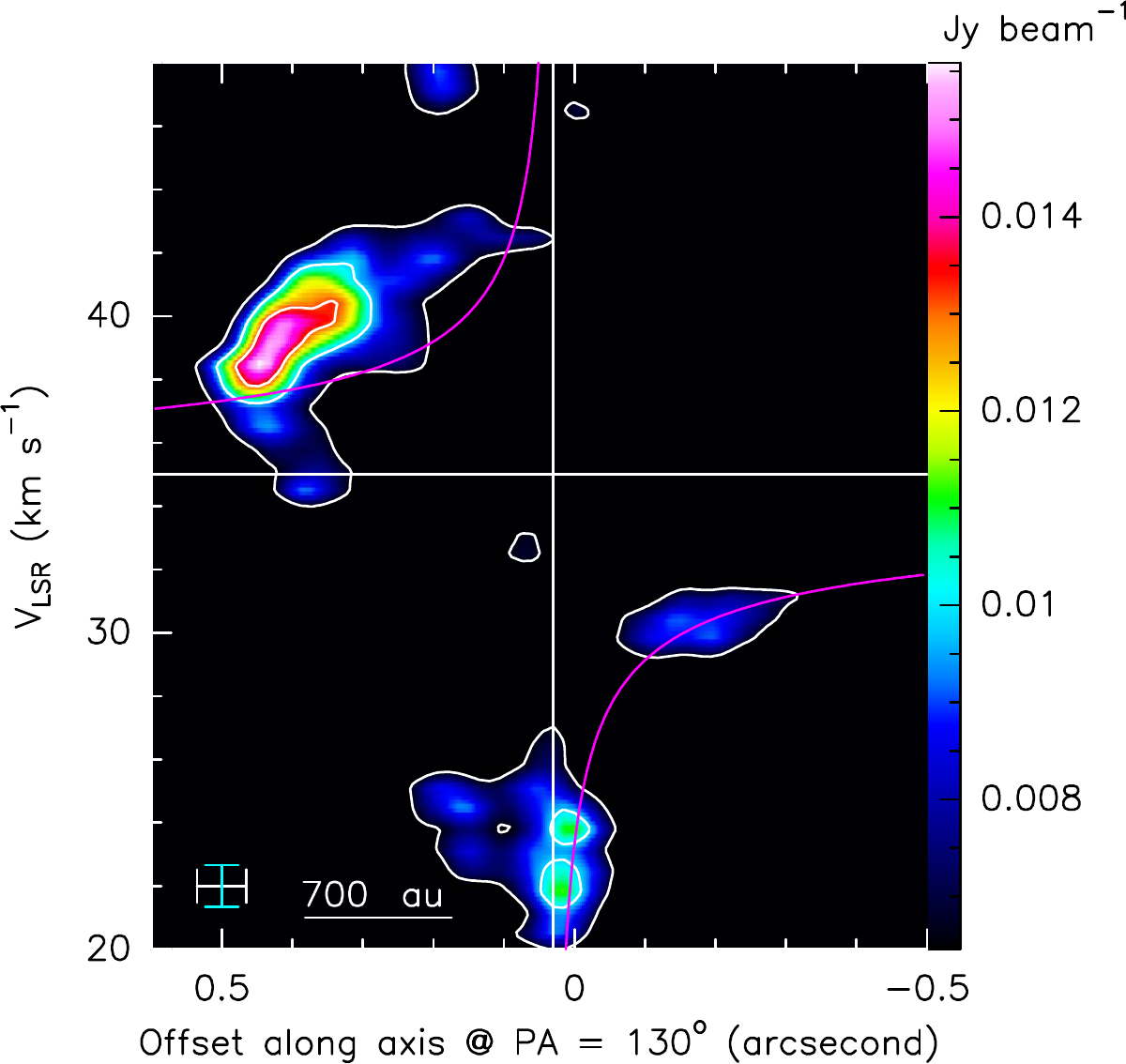}  \\
(a)  & (b) & (c) & (d) \\[6pt]
\end{tabular}
\caption{PV diagrams of CH$_3$CN $\varv=0$ $J$ = 12-11 $K$ = 3  (a), CH$_3$OH  (b),  SO$_2$ (c), and  SO (d) emission for the cut at the position of the velocity spike (see the dotted black circle in Figs. \ref{fig:M0M1}k-l). The magenta curves represent the Keplerian disk model for the central mass of 15 $M_{\odot}$. 
\label{fig:pvspur}}
\end{figure*}

There is another particular feature of the SO image that leads us to believe in the presence of a disk-wind in the source. 
The velocity field imaged with the SO molecular line data shows a velocity spike at a $\sim$0$\farcs$3 ($\sim$1000 au) separation to the NE of the central source (see the dashed black circle in Fig. \ref{fig:M0M1}l). To analyse the behaviour of the gas in this region, we present PV diagrams for a slice centred on the position of the velocity spike with PA = 130$^{\circ}$ for the tracers of extended emission (CH$_3$CN $\varv=0$ $K$=3, CH$_3$OH, SO$_2$, and SO; see Fig. \ref{fig:pvspur}). At the position of the velocity spike, mildly extended emission of CH$_3$CN $\varv=0$ as well as CH$_3$OH (for example, Figs. \ref{fig:pvspur}a and \ref{fig:pvspur}b) can be interpreted as tracing a ring structure with only the highest-velocity emission showing a hint of Keplerian rotation. In contrast, the SO$_2$ and SO profiles are consistent with Keplerian patterns (Figs. \ref{fig:pvspur}c and \ref{fig:pvspur}d) indicating Keplerian rotation of the matter at such a large distance from the central source as $\sim$1000 au. We note that the distribution of the SO emission (see the moment 0 map in Fig. \ref{fig:M0M1}k) suggests that the vicinity of the central source is opaque in this line while the velocity spike position is more optically thin and allows us to directly see this feature.

Fitting of the PV diagram for the SO emission at the velocity spike position using the KeplerFit script from \cite{Bosco2019} returned a central source mass of 15 $M_{\odot}$; this is consistent with what obtained from the disk analysis (see Table \ref{tab:disk}), suggesting a common origin. The characteristics of the SO$_2$ and SO emission can be explained assuming they trace a disk-wind in the source.
Considering a disk in Keplerian rotation, we expect the streams of the disk-wind with the highest velocities to be launched from the smallest radii; thus, the velocity spike seen in the SO Moment 1 map traces the gas coming from the innermost disk regions.

As mentioned in the introduction, one of the distinguishing parameters between the X-wind and disk-wind models is the position of the launching point within 
the disk: $<$0.1 au for X-wind \citep{Shu1995} and tens of au for disk-wind \citep{Pudritz2007}. Thus, if we want to test which of the two theories better fits our data, we need to determine the launching radius (R$_0$). For this purpose, we adopted the approach presented in \cite[their Eq. (2)]{Lee2017}, where the jet launching radius is constrained by the specific angular momentum of the jet (j), jet expansion velocity, and mass of the central object (M). 
We use our estimate of the mean specific angular momentum of $\sim$4500 au km s$^{-1}$ for the NE lobe  and of $\sim$1600 au km s$^{-1}$ for the SW lobe, as well as the mass of the protostar 15 $M_{\odot}$ (without correcting for the inclination as we did not consider it in the calculations of the angular momentum). 
The only parameter that we have not evaluated directly with our data is the outflow velocity. 
We can assume a reasonable value of $\sim$30 km s$^{-1}$ based on the 22 GHz water maser proper motions estimated in \cite{Moscadelli2019}. 
Using these estimates, we get R$_0$ $\sim$50 au for the SW lobe and $\sim$100 au for the NE lobe. 
Since the water masers typically occur in the regions where outflowing gas  collides with the ambient material and thus slows down, we can expect that the real velocity of the outflow is even higher. To take into account this possibility, we recalculated R$_0$ using the same parameters but an outflow velocity of $\sim$100 km s$^{-1}$. We get $\sim$10 au for the SW lobe and $\sim$20 au for the NE lobe.
These estimates are not consistent with the X-wind model, which requires much smaller radii, of a fraction of a au.

\begin{figure}
\centering
\begin{tabular}{cc}
  \includegraphics[width=85mm]{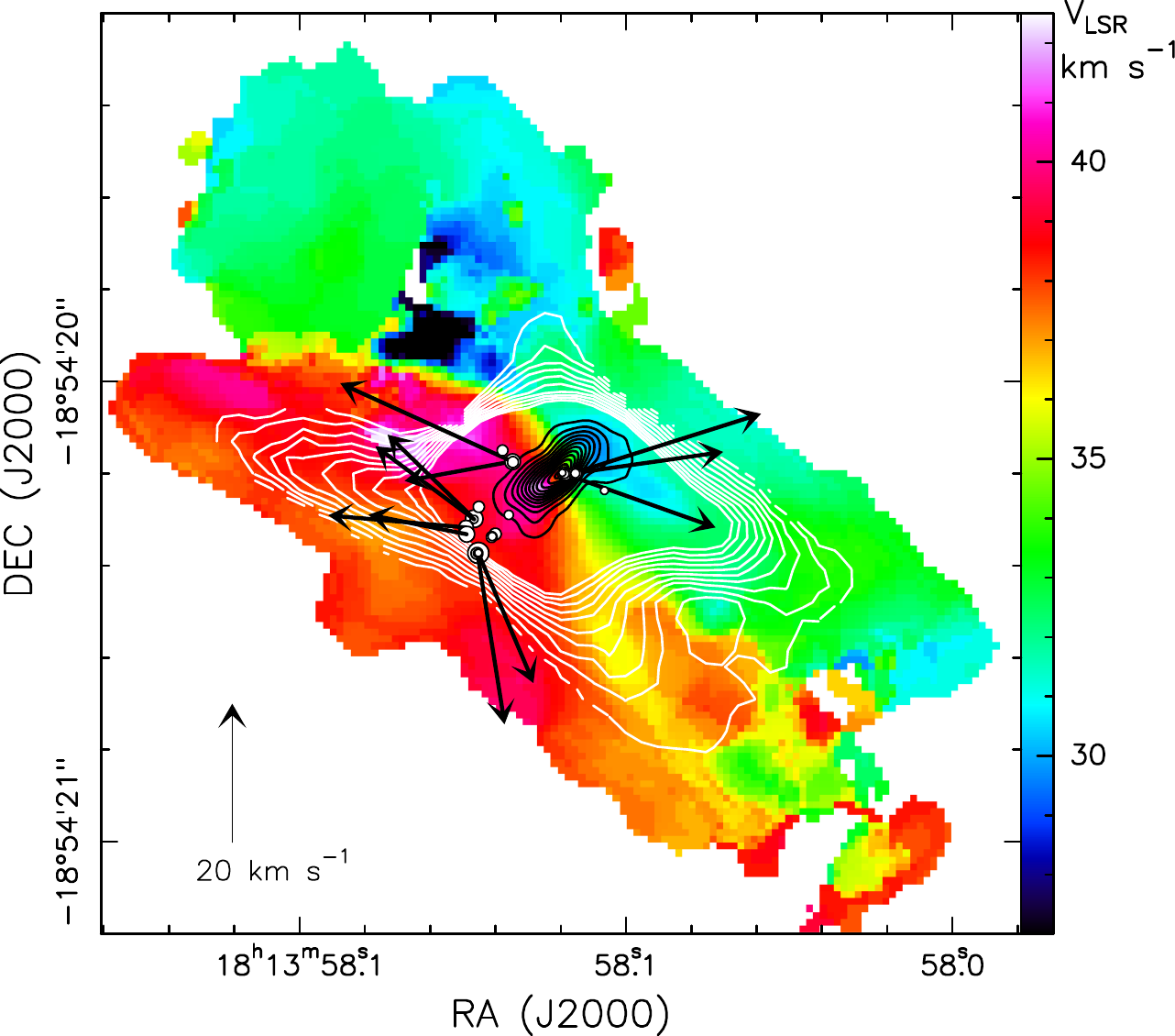} 
\end{tabular}
\caption{Velocity map of the SO emission (colour) overlaid
with the dust continuum (black contours), the SO$_2$ integrated emission (white contours), and the 22 GHz maser positions (white dots) and proper motions (black arrows).
\label{fig:masers}}
\end{figure}

A quantity of interest in disk-wind models is
the magnetic lever arm $\lambda$=R$^2_A$/R$^2_0$. Here R$_A$ is the Alfvén radius (the radius where magnetic energy density and the kinetic energy density of the outflowing material are comparable) and R$_0$ is the radius at which the field line is anchored to the disk that we previously estimated. To evaluate the magnetic lever arm for our case, we used Eq. (10) from \cite{Ferreira2006} and the estimated values of the  specific angular momentum, outflow propagation velocity, and protostar mass. For a velocity of the outflow  of $\sim$30 km s$^{-1}$, we obtain $\lambda$ of $\sim$2.4 and $\sim$4, and R$_A$ of $\sim$80 and $\sim$200 for the SW and NE lobe, respectively. Such $\lambda$ are in good agreement with the typical values (between 2-10) expected for disk-winds in massive protostars.

Finally, the image in Fig. \ref{fig:masers}, combining the SO, SO$_2$, and 1.3 mm continuum emission obtained in this work, as well as the 22 GHz water masers from \cite{Moscadelli2019}, shows that all the outflow tracers originate from a common large rotating structure but at different scales.
Figure \ref{fig:masers} confirms that the proper motions of the 22 GHz water masers correlate well with the velocity field of the molecular data, suggesting that each of the maser groups co-propagates with the outflow material.
As we mentioned above, in the case of massive protostars, high-resolution and high-sensitivity observations of masers are required to resolve the launching region of a disk-wind and  confirm its nature. 
In G11.92-0.61 MM1a, the 22 GHz water masers trace the launching base of the outflow but, unfortunately, not the streamlines of a disk-wind. 
In conclusion, G11.92-0.61 MM1a seems to be a unique example of a massive protostar in which bright molecular outflow emission (1) reveals clear signs of rotation over  $\sim$3400 au from the central source, and (2) shows strong indications for the presence of a disk-wind launched from small radii ($\sim$50-100 au) of the disk. Our study demonstrates that high-resolution molecular line data can  suffice to reveal disk-winds in massive protostars.

\section{Conclusions}

 We have carried out an ALMA Band 6 observation of the massive protostar G11.92-0.61 MM1. The main results  can be summarised as follows:

   \begin{enumerate}
      \item We confirm the existence of a disk around the source MM1a. The estimate of the disk diameter depends on the tracer, ranging from 500 au in the millimetre continuum to 700 au in the CH$_3$CN and CH$_3$OH lines.
      \item The molecular line data reveal the presence of two distinct velocity fields within the disk: (1) Keplerian rotation at radii of $<$300 au, traced by high-velocity CH$_3$CN emission, and (2) sub-Keplerian rotation at larger disk radii, $>$300 au, traced by mid-velocity CH$_3$CN   emission. 
      \item Keplerian fitting of the molecular data associated with the disk gives a central source mass of 20 $M_{\odot}$, which is about two times lower than the previous estimates for MM1a and is consistent with  the source luminosity of $\sim$10$^4$ L$_{\rm sun}$. 
      \item The source MM1b appears to be a distinct source rather than a disk fragment as no molecular emission associated with MM1b and no connection between MM1a and MM1b are detected.    
      \item A strong collimated SO$_2$ and SO outflow from MM1a is probed up to $\sim$3400 au from MM1a.  
      \item The observed velocity pattern of the SO$_2$ and SO emission indicates that the outflow is rotating about its axis. 
      A decrease in the rotation velocity of $\sim$1.5 km s$^{-1}$ over $\sim$1500 au is found in both lobes of the outflow. Along the lobes, the specific angular momentum is approximately constant: $\sim$4720 au km s$^{-1}$ for the NE lobe and $\sim$1420 au km s$^{-1}$ for the SW lobe.
      \item The Keplerian rotation profile obtained for the molecular outflow tracers suggests magneto-centrifugal launching of the wind. Considering the estimated specific angular momentum and mass of the central object, we derive a launching radius of the outflow of $\sim$50-100 au, which is consistent with the disk-wind model, while the X-wind model requires much smaller launching radii. 
      \item Comparison of the newly acquired molecular data and 22 GHz water maser proper motions indicates that the water masers trace the disk-wind interaction region.
   \end{enumerate}

\begin{acknowledgements}
This paper makes use of the following ALMA data: ADS/JAO.ALMA$\#$2019.1.01639.S. ALMA is a partnership of ESO (representing its member states), NSF (USA) and NINS (Japan), together with NRC (Canada), MOST and ASIAA (Taiwan), and KASI (Republic of Korea), in cooperation with the Republic of Chile. The Joint ALMA Observatory is operated by ESO, AUI/NRAO and NAOJ.
O.B. acknowledges financial support from the Italian Ministry of University and Research - Project Proposal CIR01$\_$00010.
\end{acknowledgements}

\bibliographystyle{aa}

  \bibliography{bib}

\begin{thebibliography}{46}
\expandafter\ifx\csname natexlab\endcsname\relax\def\natexlab#1{#1}\fi

\bibitem[{{Bally} \& {Lada}(1983)}]{Bally1983}
{Bally}, J. \& {Lada}, C.~J. 1983, \apj, 265, 824

\bibitem[{{Blandford} \& {Payne}(1982)}]{Blandford1982}
{Blandford}, R.~D. \& {Payne}, D.~G. 1982, \mnras, 199, 883

\bibitem[{{Bosco} {et~al.}(2019){Bosco}, {Beuther}, {Ahmadi}, {Mottram}, {Kuiper}, {Linz}, {Maud}, {Winters}, {Henning}, {Feng}, {Peters}, {Semenov}, {Klaassen}, {Schilke}, {Urquhart}, {Beltr{\'a}n}, {Lumsden}, {Leurini}, {Moscadelli}, {Cesaroni}, {S{\'a}nchez-Monge}, {Palau}, {Pudritz}, {Wyrowski}, \& {Longmore}}]{Bosco2019}
{Bosco}, F., {Beuther}, H., {Ahmadi}, A., {et~al.} 2019, \aap, 629, A10

\bibitem[{{Breen} \& {Ellingsen}(2011)}]{Breen2011}
{Breen}, S.~L. \& {Ellingsen}, S.~P. 2011, \mnras, 416, 178

\bibitem[{{Canto}(1980)}]{Canto1980}
{Canto}, J. 1980, \aap, 86, 327

\bibitem[{{CASA Team} {et~al.}(2022){CASA Team}, {Bean}, {Bhatnagar}, {Castro}, {Donovan Meyer}, {Emonts}, {Garcia}, {Garwood}, {Golap}, {Gonzalez Villalba}, {Harris}, {Hayashi}, {Hoskins}, {Hsieh}, {Jagannathan}, {Kawasaki}, {Keimpema}, {Kettenis}, {Lopez}, {Marvil}, {Masters}, {McNichols}, {Mehringer}, {Miel}, {Moellenbrock}, {Montesino}, {Nakazato}, {Ott}, {Petry}, {Pokorny}, {Raba}, {Rau}, {Schiebel}, {Schweighart}, {Sekhar}, {Shimada}, {Small}, {Steeb}, {Sugimoto}, {Suoranta}, {Tsutsumi}, {van Bemmel}, {Verkouter}, {Wells}, {Xiong}, {Szomoru}, {Griffith}, {Glendenning}, \& {Kern}}]{CASA2022}
{CASA Team}, {Bean}, B., {Bhatnagar}, S., {et~al.} 2022, \pasp, 134, 114501

\bibitem[{{Comrie} {et~al.}(2020){Comrie}, {Pi{\'n}ska}, {Simmonds}, \& {Taylor}}]{Comrie2020}
{Comrie}, A., {Pi{\'n}ska}, A., {Simmonds}, R., \& {Taylor}, A.~R. 2020, Astronomy and Computing, 32, 100389

\bibitem[{{Cyganowski} {et~al.}(2009){Cyganowski}, {Brogan}, {Hunter}, \& {Churchwell}}]{Cyganowski2009}
{Cyganowski}, C.~J., {Brogan}, C.~L., {Hunter}, T.~R., \& {Churchwell}, E. 2009, \apj, 702, 1615

\bibitem[{{Cyganowski} {et~al.}(2011{\natexlab{a}}){Cyganowski}, {Brogan}, {Hunter}, \& {Churchwell}}]{Cyganowski2011b}
{Cyganowski}, C.~J., {Brogan}, C.~L., {Hunter}, T.~R., \& {Churchwell}, E. 2011{\natexlab{a}}, \apj, 743, 56

\bibitem[{{Cyganowski} {et~al.}(2011{\natexlab{b}}){Cyganowski}, {Brogan}, {Hunter}, {Churchwell}, \& {Zhang}}]{Cyganowski2011}
{Cyganowski}, C.~J., {Brogan}, C.~L., {Hunter}, T.~R., {Churchwell}, E., \& {Zhang}, Q. 2011{\natexlab{b}}, \apj, 729, 124

\bibitem[{{Cyganowski} {et~al.}(2017){Cyganowski}, {Brogan}, {Hunter}, {Smith}, {Kruijssen}, {Bonnell}, \& {Zhang}}]{Cyganowski2017}
{Cyganowski}, C.~J., {Brogan}, C.~L., {Hunter}, T.~R., {et~al.} 2017, \mnras, 468, 3694

\bibitem[{{Cyganowski} {et~al.}(2008){Cyganowski}, {Whitney}, {Holden}, {Braden}, {Brogan}, {Churchwell}, {Indebetouw}, {Watson}, {Babler}, {Benjamin}, {Gomez}, {Meade}, {Povich}, {Robitaille}, \& {Watson}}]{Cyganowski2008}
{Cyganowski}, C.~J., {Whitney}, B.~A., {Holden}, E., {et~al.} 2008, \aj, 136, 2391

\bibitem[{{Ferreira} {et~al.}(2006){Ferreira}, {Dougados}, \& {Cabrit}}]{Ferreira2006}
{Ferreira}, J., {Dougados}, C., \& {Cabrit}, S. 2006, \aap, 453, 785

\bibitem[{{Frank} {et~al.}(2014){Frank}, {Ray}, {Cabrit}, {Hartigan}, {Arce}, {Bacciotti}, {Bally}, {Benisty}, {Eisl{\"o}ffel}, {G{\"u}del}, {Lebedev}, {Nisini}, \& {Raga}}]{Frank2014}
{Frank}, A., {Ray}, T.~P., {Cabrit}, S., {et~al.} 2014, in Protostars and Planets VI, ed. H.~{Beuther}, R.~S. {Klessen}, C.~P. {Dullemond}, \& T.~{Henning}, 451--474

\bibitem[{{Gieser} {et~al.}(2021){Gieser}, {Beuther}, {Semenov}, {Ahmadi}, {Suri}, {M{\"o}ller}, {Beltr{\'a}n}, {Klaassen}, {Zhang}, {Urquhart}, {Henning}, {Feng}, {Galv{\'a}n-Madrid}, {de Souza Magalh{\~a}es}, {Moscadelli}, {Longmore}, {Leurini}, {Kuiper}, {Peters}, {Menten}, {Csengeri}, {Fuller}, {Wyrowski}, {Lumsden}, {S{\'a}nchez-Monge}, {Maud}, {Linz}, {Palau}, {Schilke}, {Pety}, {Pudritz}, {Winters}, \& {Pi{\'e}tu}}]{Gieser2021}
{Gieser}, C., {Beuther}, H., {Semenov}, D., {et~al.} 2021, \aap, 648, A66

\bibitem[{{Goddi} {et~al.}(2020){Goddi}, {Ginsburg}, {Maud}, {Zhang}, \& {Zapata}}]{Goddi2020}
{Goddi}, C., {Ginsburg}, A., {Maud}, L.~T., {Zhang}, Q., \& {Zapata}, L.~A. 2020, \apj, 905, 25

\bibitem[{{Hildebrand}(1983)}]{Hildebrand1983}
{Hildebrand}, R.~H. 1983, \qjras, 24, 267

\bibitem[{{Hirota} {et~al.}(2017){Hirota}, {Machida}, {Matsushita}, {Motogi}, {Matsumoto}, {Kim}, {Burns}, \& {Honma}}]{Hirota2017}
{Hirota}, T., {Machida}, M.~N., {Matsushita}, Y., {et~al.} 2017, Nature Astronomy, 1, 0146

\bibitem[{{Hofner} \& {Churchwell}(1996)}]{Hofner1996}
{Hofner}, P. \& {Churchwell}, E. 1996, \aaps, 120, 283

\bibitem[{{Ilee} {et~al.}(2018){Ilee}, {Cyganowski}, {Brogan}, {Hunter}, {Forgan}, {Haworth}, {Clarke}, \& {Harries}}]{Ilee2018}
{Ilee}, J.~D., {Cyganowski}, C.~J., {Brogan}, C.~L., {et~al.} 2018, \apjl, 869, L24

\bibitem[{{Ilee} {et~al.}(2016){Ilee}, {Cyganowski}, {Nazari}, {Hunter}, {Brogan}, {Forgan}, \& {Zhang}}]{Ilee2016}
{Ilee}, J.~D., {Cyganowski}, C.~J., {Nazari}, P., {et~al.} 2016, \mnras, 462, 4386

\bibitem[{{Kim} {et~al.}(2008){Kim}, {Hirota}, {Honma}, {Kobayashi}, {Bushimata}, {Choi}, {Imai}, {Iwadate}, {Jike}, {Kameno}, {Kameya}, {Kamohara}, {Kan-Ya}, {Kawaguchi}, {Kuji}, {Kurayama}, {Manabe}, {Matsui}, {Matsumoto}, {Miyaji}, {Nagayama}, {Nakagawa}, {Oh}, {Omodaka}, {Oyama}, {Sakai}, {Sasao}, {Sato}, {Sato}, {Shibata}, {Tamura}, \& {Yamashita}}]{Kim2008}
{Kim}, M.~K., {Hirota}, T., {Honma}, M., {et~al.} 2008, \pasj, 60, 991

\bibitem[{{Kuiper} {et~al.}(2011){Kuiper}, {Klahr}, {Beuther}, \& {Henning}}]{Kuiper2011}
{Kuiper}, R., {Klahr}, H., {Beuther}, H., \& {Henning}, T. 2011, \apj, 732, 20

\bibitem[{{Lee} {et~al.}(2017){Lee}, {Ho}, {Li}, {Hirano}, {Zhang}, \& {Shang}}]{Lee2017}
{Lee}, C.-F., {Ho}, P. T.~P., {Li}, Z.-Y., {et~al.} 2017, Nature Astronomy, 1, 0152

\bibitem[{{Mart{\'\i}n} {et~al.}(2019){Mart{\'\i}n}, {Mart{\'\i}n-Pintado}, {Blanco-S{\'a}nchez}, {Rivilla}, {Rodr{\'\i}guez-Franco}, \& {Rico-Villas}}]{Martin2019}
{Mart{\'\i}n}, S., {Mart{\'\i}n-Pintado}, J., {Blanco-S{\'a}nchez}, C., {et~al.} 2019, \aap, 631, A159

\bibitem[{{Matthews} {et~al.}(2010){Matthews}, {Greenhill}, {Goddi}, {Chandler}, {Humphreys}, \& {Kunz}}]{Matthews2010}
{Matthews}, L.~D., {Greenhill}, L.~J., {Goddi}, C., {et~al.} 2010, \apj, 708, 80

\bibitem[{{McCaughrean} {et~al.}(1994){McCaughrean}, {Rayner}, \& {Zinnecker}}]{McCaughrean1994}
{McCaughrean}, M.~J., {Rayner}, J.~T., \& {Zinnecker}, H. 1994, \apjl, 436, L189

\bibitem[{{Miyawaki} {et~al.}(2022){Miyawaki}, {Hayashi}, \& {Hasegawa}}]{Miyawaki2022}
{Miyawaki}, R., {Hayashi}, M., \& {Hasegawa}, T. 2022, \pasj, 74, 705

\bibitem[{{Moscadelli} {et~al.}(2011){Moscadelli}, {Cesaroni}, {Rioja}, {Dodson}, \& {Reid}}]{Moscadelli2011}
{Moscadelli}, L., {Cesaroni}, R., {Rioja}, M.~J., {Dodson}, R., \& {Reid}, M.~J. 2011, \aap, 526, A66

\bibitem[{{Moscadelli} {et~al.}(2016){Moscadelli}, {S{\'a}nchez-Monge}, {Goddi}, {Li}, {Sanna}, {Cesaroni}, {Pestalozzi}, {Molinari}, \& {Reid}}]{Moscadelli2016}
{Moscadelli}, L., {S{\'a}nchez-Monge}, {\'A}., {Goddi}, C., {et~al.} 2016, \aap, 585, A71

\bibitem[{{Moscadelli} {et~al.}(2022){Moscadelli}, {Sanna}, {Beuther}, {Oliva}, \& {Kuiper}}]{Moscadelli2022}
{Moscadelli}, L., {Sanna}, A., {Beuther}, H., {Oliva}, A., \& {Kuiper}, R. 2022, Nature Astronomy, 6, 1068

\bibitem[{{Moscadelli} {et~al.}(2019){Moscadelli}, {Sanna}, {Goddi}, {Krishnan}, {Massi}, \& {Bacciotti}}]{Moscadelli2019}
{Moscadelli}, L., {Sanna}, A., {Goddi}, C., {et~al.} 2019, \aap, 631, A74

\bibitem[{{Pascucci} {et~al.}(2023){Pascucci}, {Cabrit}, {Edwards}, {Gorti}, {Gressel}, \& {Suzuki}}]{Pascucci2023}
{Pascucci}, I., {Cabrit}, S., {Edwards}, S., {et~al.} 2023, in Astronomical Society of the Pacific Conference Series, Vol. 534, Protostars and Planets VII, ed. S.~{Inutsuka}, Y.~{Aikawa}, T.~{Muto}, K.~{Tomida}, \& M.~{Tamura}, 567

\bibitem[{{P{\'e}rez} {et~al.}(2012){P{\'e}rez}, {Carpenter}, {Chandler}, {Isella}, {Andrews}, {Ricci}, {Calvet}, {Corder}, {Deller}, {Dullemond}, {Greaves}, {Harris}, {Henning}, {Kwon}, {Lazio}, {Linz}, {Mundy}, {Sargent}, {Storm}, {Testi}, \& {Wilner}}]{Perez2012}
{P{\'e}rez}, L.~M., {Carpenter}, J.~M., {Chandler}, C.~J., {et~al.} 2012, \apjl, 760, L17

\bibitem[{{Pudritz} {et~al.}(2007){Pudritz}, {Ouyed}, {Fendt}, \& {Brandenburg}}]{Pudritz2007}
{Pudritz}, R.~E., {Ouyed}, R., {Fendt}, C., \& {Brandenburg}, A. 2007, in Protostars and Planets V, ed. B.~{Reipurth}, D.~{Jewitt}, \& K.~{Keil}, 277

\bibitem[{{S{\'a}nchez-Monge} {et~al.}(2013){S{\'a}nchez-Monge}, {Cesaroni}, {Beltr{\'a}n}, {Kumar}, {Stanke}, {Zinnecker}, {Etoka}, {Galli}, {Hummel}, {Moscadelli}, {Preibisch}, {Ratzka}, {van der Tak}, {Vig}, {Walmsley}, \& {Wang}}]{Sanchez2013}
{S{\'a}nchez-Monge}, {\'A}., {Cesaroni}, R., {Beltr{\'a}n}, M.~T., {et~al.} 2013, \aap, 552, L10

\bibitem[{{Sanna} {et~al.}(2019){Sanna}, {K{\"o}lligan}, {Moscadelli}, {Kuiper}, {Cesaroni}, {Pillai}, {Menten}, {Zhang}, {Caratti o Garatti}, {Goddi}, {Leurini}, \& {Carrasco-Gonz{\'a}lez}}]{Sanna2019}
{Sanna}, A., {K{\"o}lligan}, A., {Moscadelli}, L., {et~al.} 2019, \aap, 623, A77

\bibitem[{{Sanna} {et~al.}(2018){Sanna}, {Moscadelli}, {Goddi}, {Krishnan}, \& {Massi}}]{Sanna2018}
{Sanna}, A., {Moscadelli}, L., {Goddi}, C., {Krishnan}, V., \& {Massi}, F. 2018, \aap, 619, A107

\bibitem[{{Sanna} {et~al.}(2012){Sanna}, {Reid}, {Carrasco-Gonz{\'a}lez}, {Menten}, {Brunthaler}, {Moscadelli}, \& {Rygl}}]{Sanna2012}
{Sanna}, A., {Reid}, M.~J., {Carrasco-Gonz{\'a}lez}, C., {et~al.} 2012, \apj, 745, 191

\bibitem[{{Sato} {et~al.}(2014){Sato}, {Wu}, {Immer}, {Zhang}, {Sanna}, {Reid}, {Dame}, {Brunthaler}, \& {Menten}}]{Sato2014}
{Sato}, M., {Wu}, Y.~W., {Immer}, K., {et~al.} 2014, \apj, 793, 72

\bibitem[{{Schuller} {et~al.}(2009){Schuller}, {Menten}, {Contreras}, {Wyrowski}, {Schilke}, {Bronfman}, {Henning}, {Walmsley}, {Beuther}, {Bontemps}, {Cesaroni}, {Deharveng}, {Garay}, {Herpin}, {Lefloch}, {Linz}, {Mardones}, {Minier}, {Molinari}, {Motte}, {Nyman}, {Reveret}, {Risacher}, {Russeil}, {Schneider}, {Testi}, {Troost}, {Vasyunina}, {Wienen}, {Zavagno}, {Kovacs}, {Kreysa}, {Siringo}, \& {Wei{\ss}}}]{Schuller2009}
{Schuller}, F., {Menten}, K.~M., {Contreras}, Y., {et~al.} 2009, \aap, 504, 415

\bibitem[{{Seifried} {et~al.}(2011){Seifried}, {Banerjee}, {Klessen}, {Duffin}, \& {Pudritz}}]{Seifried2011}
{Seifried}, D., {Banerjee}, R., {Klessen}, R.~S., {Duffin}, D., \& {Pudritz}, R.~E. 2011, \mnras, 417, 1054

\bibitem[{{Shu} {et~al.}(1995){Shu}, {Najita}, {Ostriker}, \& {Shang}}]{Shu1995}
{Shu}, F.~H., {Najita}, J., {Ostriker}, E.~C., \& {Shang}, H. 1995, \apjl, 455, L155

\bibitem[{{Suriano} {et~al.}(2019){Suriano}, {Li}, {Krasnopolsky}, {Suzuki}, \& {Shang}}]{Suriano2019}
{Suriano}, S.~S., {Li}, Z.-Y., {Krasnopolsky}, R., {Suzuki}, T.~K., \& {Shang}, H. 2019, \mnras, 484, 107

\bibitem[{{Vorobyov} {et~al.}(2018){Vorobyov}, {Akimkin}, {Stoyanovskaya}, {Pavlyuchenkov}, \& {Liu}}]{Vorobyov2018}
{Vorobyov}, E.~I., {Akimkin}, V., {Stoyanovskaya}, O., {Pavlyuchenkov}, Y., \& {Liu}, H.~B. 2018, \aap, 614, A98

\bibitem[{{Zhang} {et~al.}(2018){Zhang}, {Higuchi}, {Sakai}, {Oya}, {L{\'o}pez-Sepulcre}, {Imai}, {Sakai}, {Watanabe}, {Ceccarelli}, {Lefloch}, \& {Yamamoto}}]{Zhang2018}
{Zhang}, Y., {Higuchi}, A.~E., {Sakai}, N., {et~al.} 2018, \apj, 864, 76

\end{thebibliography}

\begin{appendix}
\onecolumn
\section{Position--velocity diagrams perpendicular to the outflow axis}
\begin{figure}[h]
\centering
\begin{tabular}{ccc}
  \includegraphics[width=60mm]{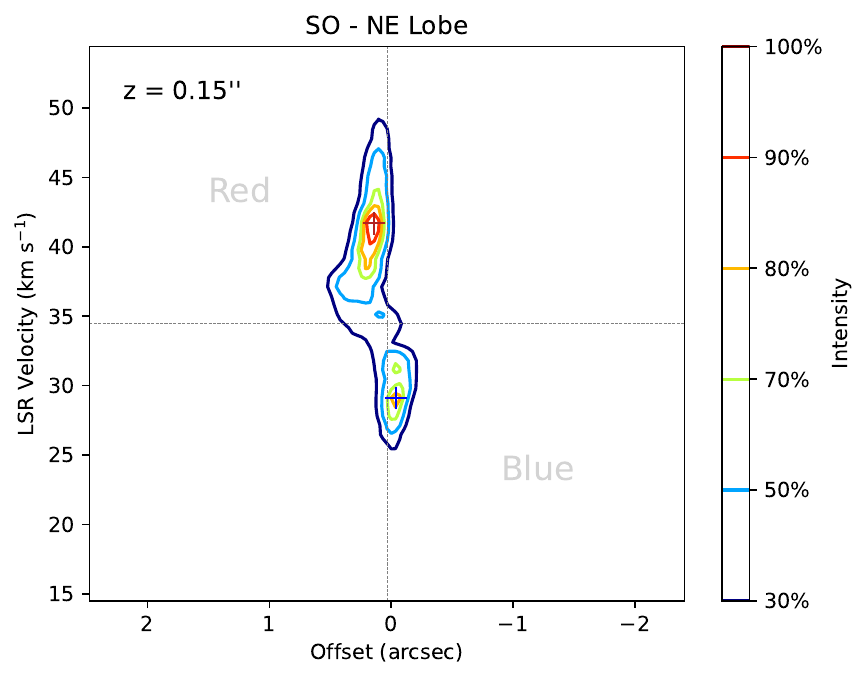}  &
  \includegraphics[width=60mm]{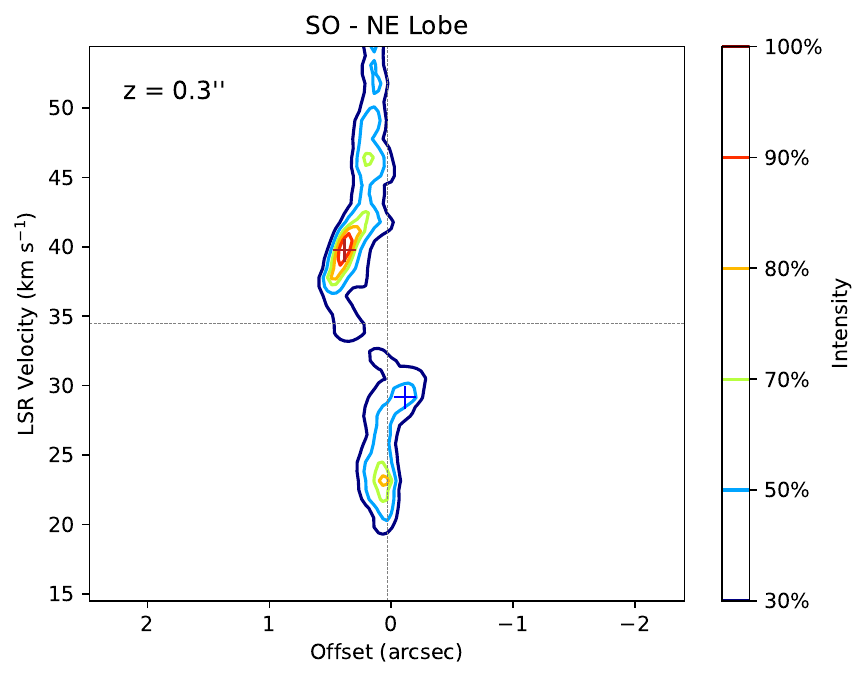}  &
  \includegraphics[width=60mm]{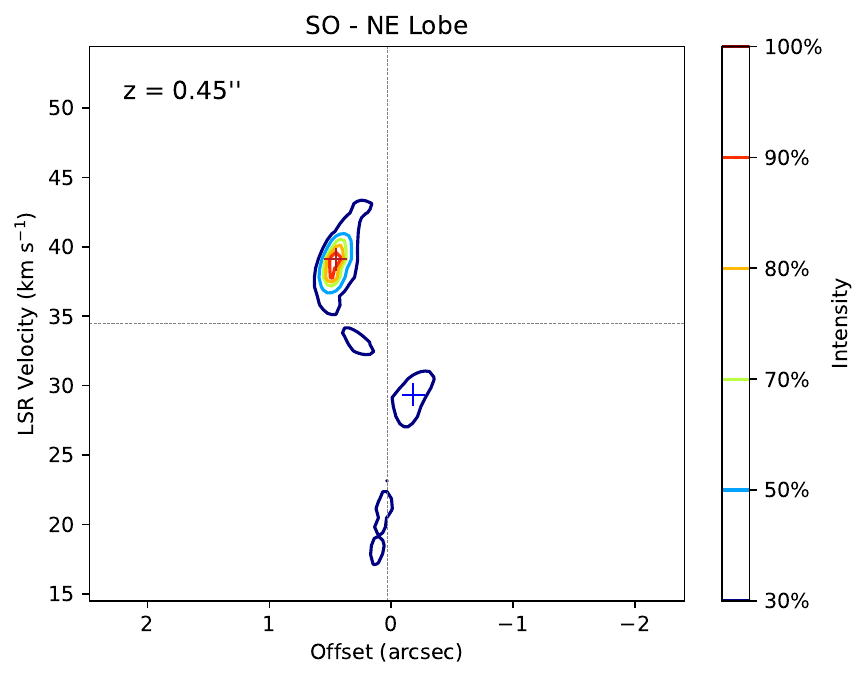}  \\
   \includegraphics[width=60mm]{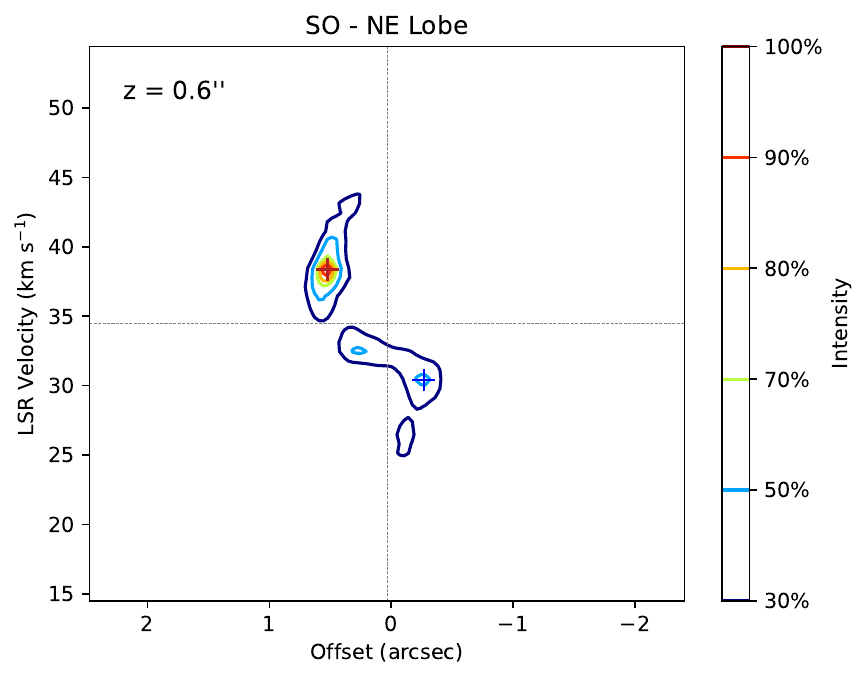}  &
  \includegraphics[width=60mm]{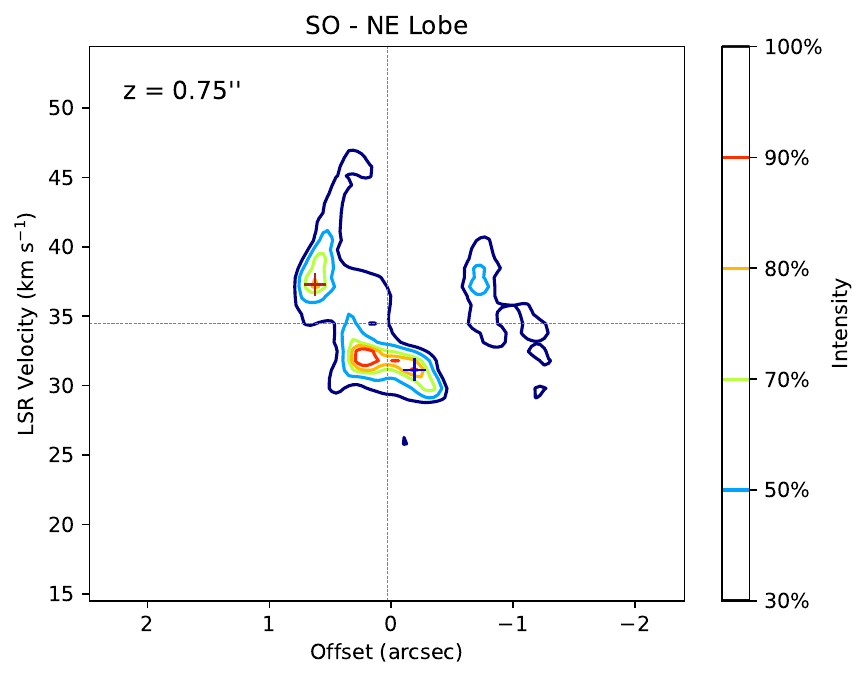}  &
\end{tabular}
\caption{PV diagrams for the cuts with steps of 0$\farcs$15 through the NE lobe of the SO-traced outflow. The offset of the cut is indicated in the top-left corner of each panel. The red (blue) cross indicates the peak in the red (blue) part of each diagram. 
The vertical and horizontal lines mark, respectively, the position of the protostar and the systemic velocity, V$_{\rm star}$, obtained from the Keplerian fit (see Table \ref{tab:disk}). The zero offset corresponds to the fitted position of the 1.3 mm continuum peak.}
\label{fig:pvsone}
\end{figure}

\begin{figure*}
\centering
\begin{tabular}{ccc}
  \includegraphics[width=60mm]{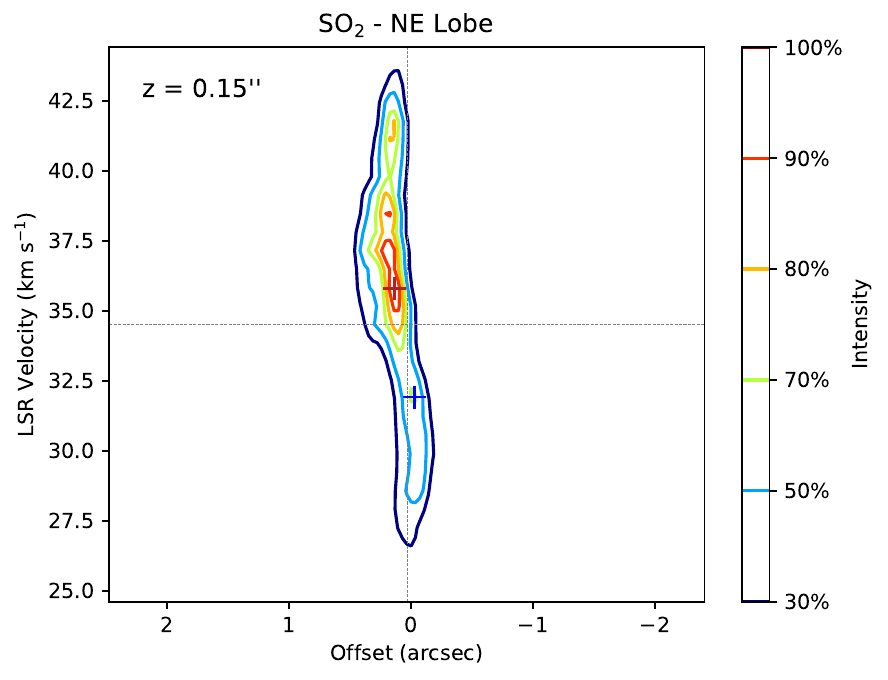}  &
  \includegraphics[width=60mm]{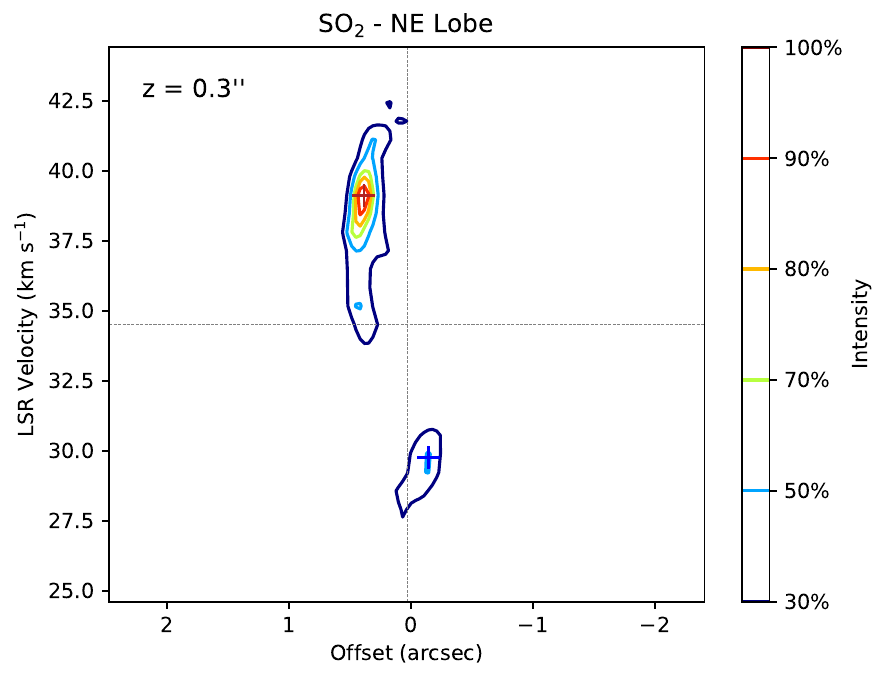}  &
  \includegraphics[width=60mm]{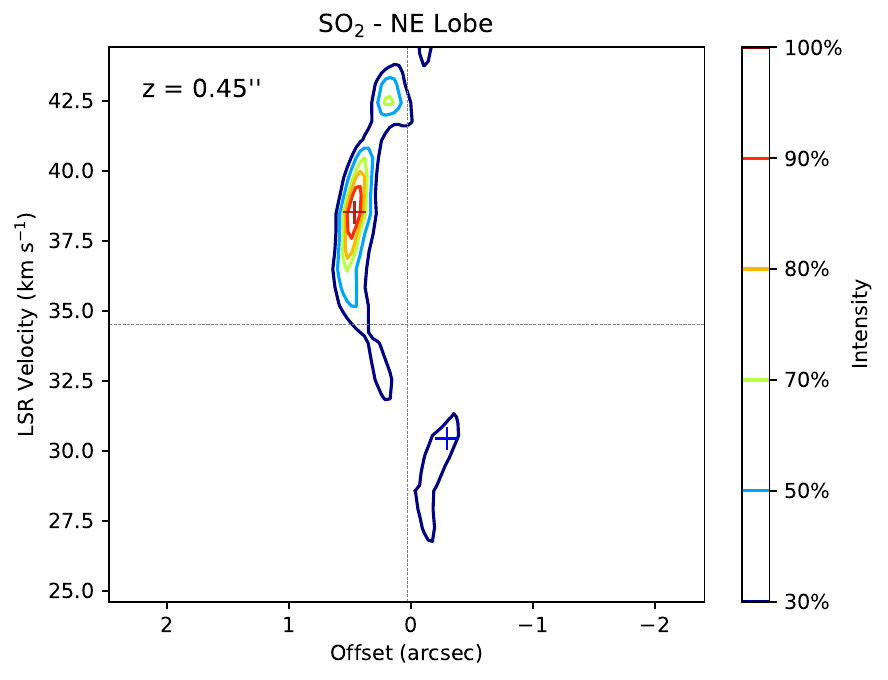}  \\
   \includegraphics[width=60mm]{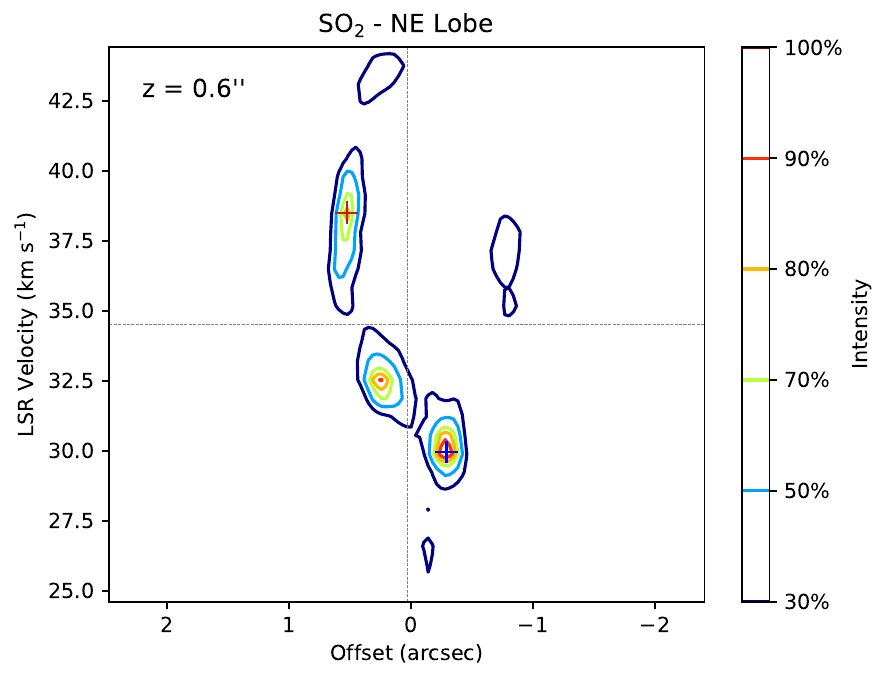}  &
  \includegraphics[width=60mm]{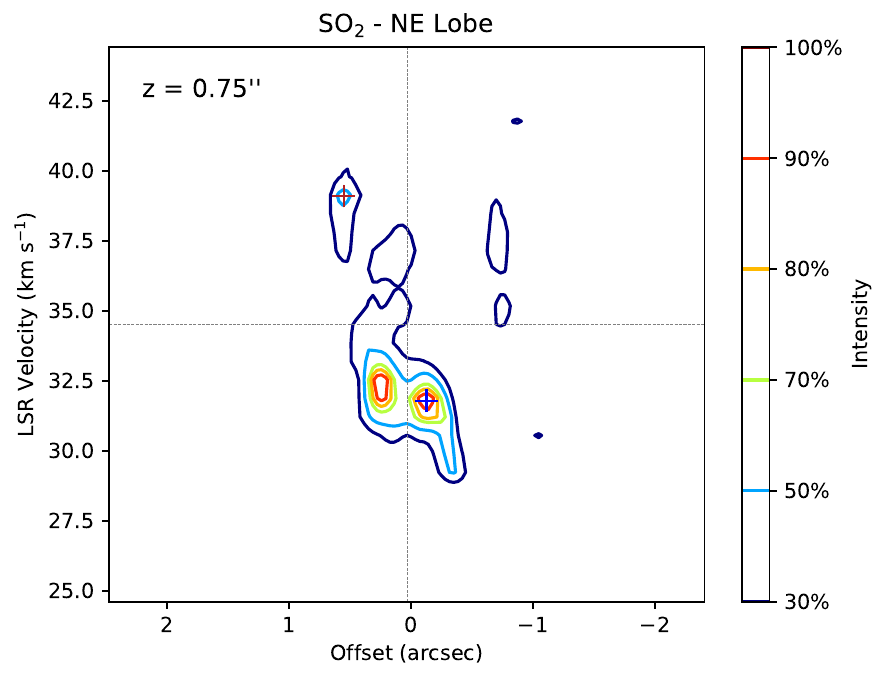}  &
\end{tabular}
\caption{Same as Fig. A.1 but for the SO$_2$ outflow.}
\label{fig:pvso2ne}
\end{figure*}

\clearpage
\begin{figure*}
\centering
\begin{tabular}{ccc}
  \includegraphics[width=60mm]{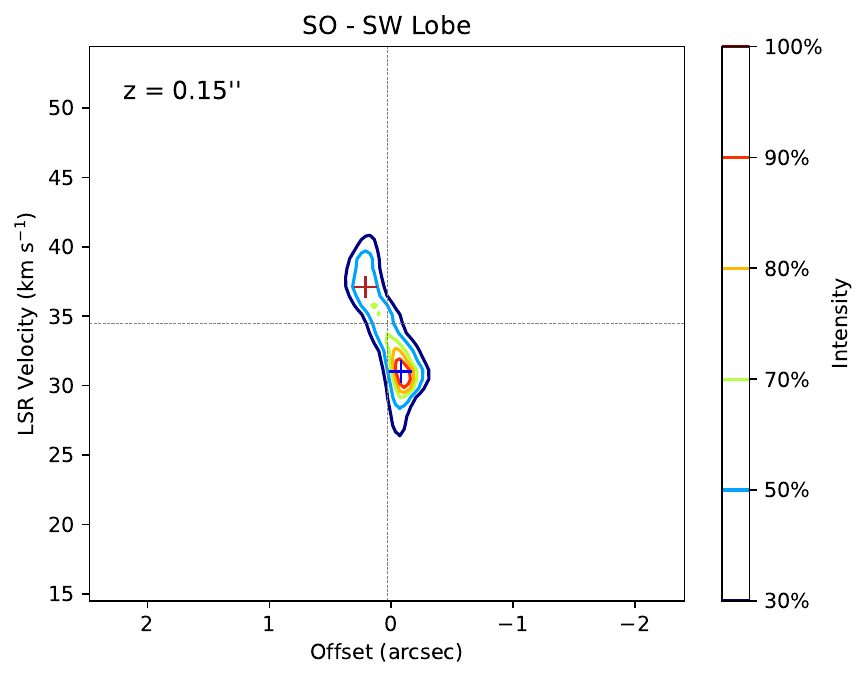}  &
  \includegraphics[width=60mm]{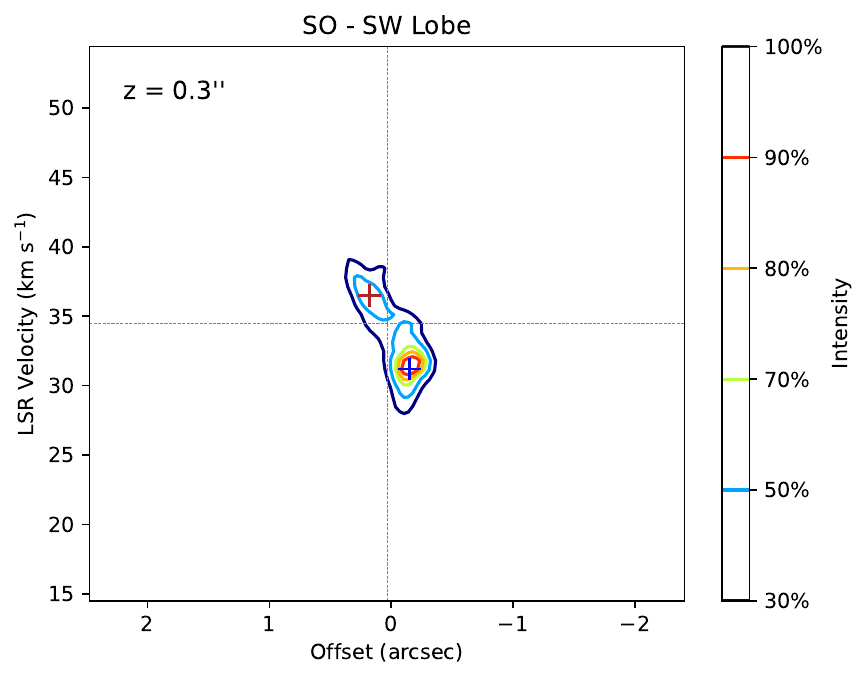}  &
  \includegraphics[width=60mm]{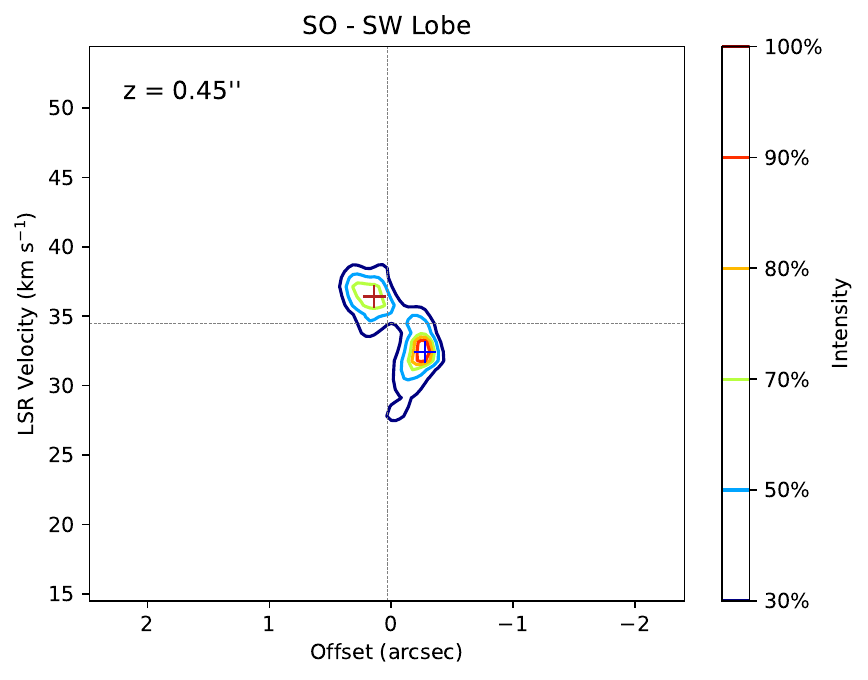}  \\
   \includegraphics[width=60mm]{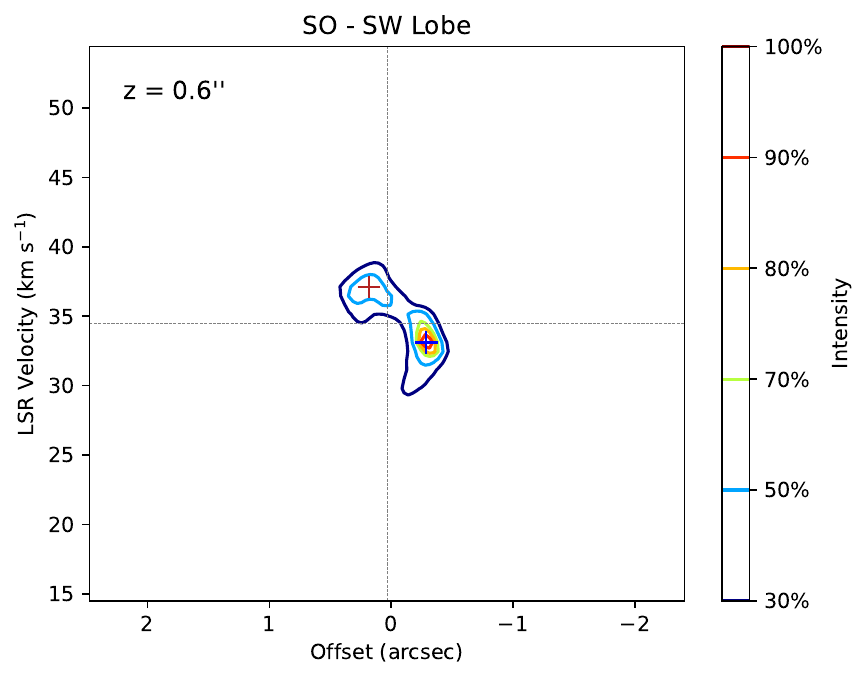}  &
  \includegraphics[width=60mm]{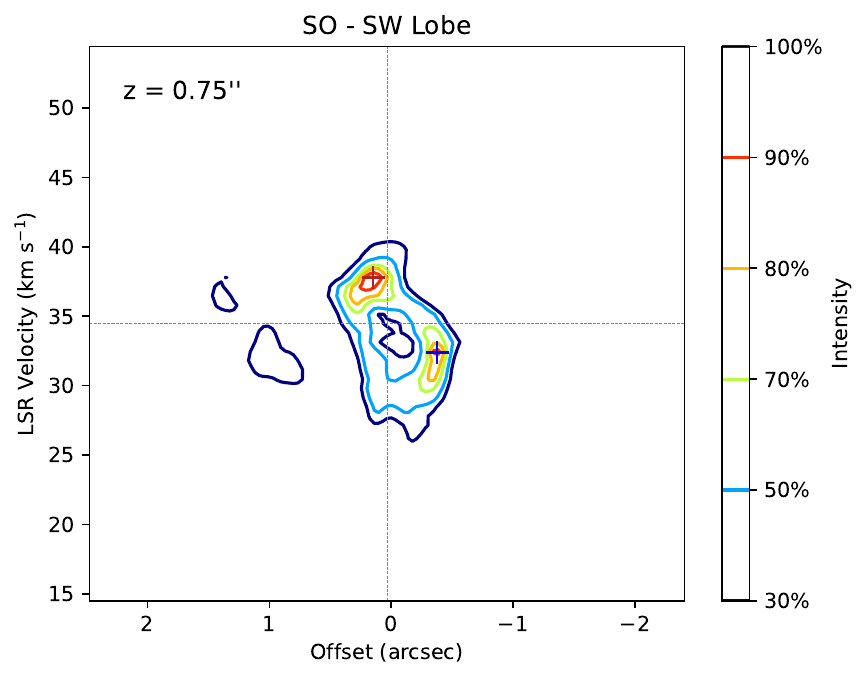}  &
\end{tabular}
\caption{Same as Fig. A.1 but for the SW lobe.
\label{fig:pvsosw}}
\end{figure*}

\begin{figure*}
\centering
\begin{tabular}{ccc}
  \includegraphics[width=60mm]{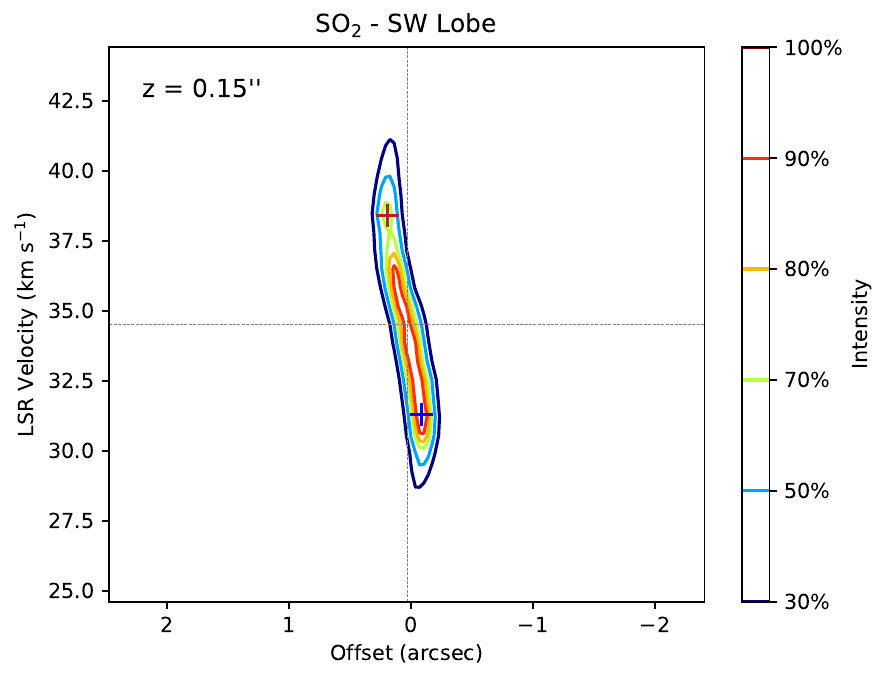}  &
  \includegraphics[width=60mm]{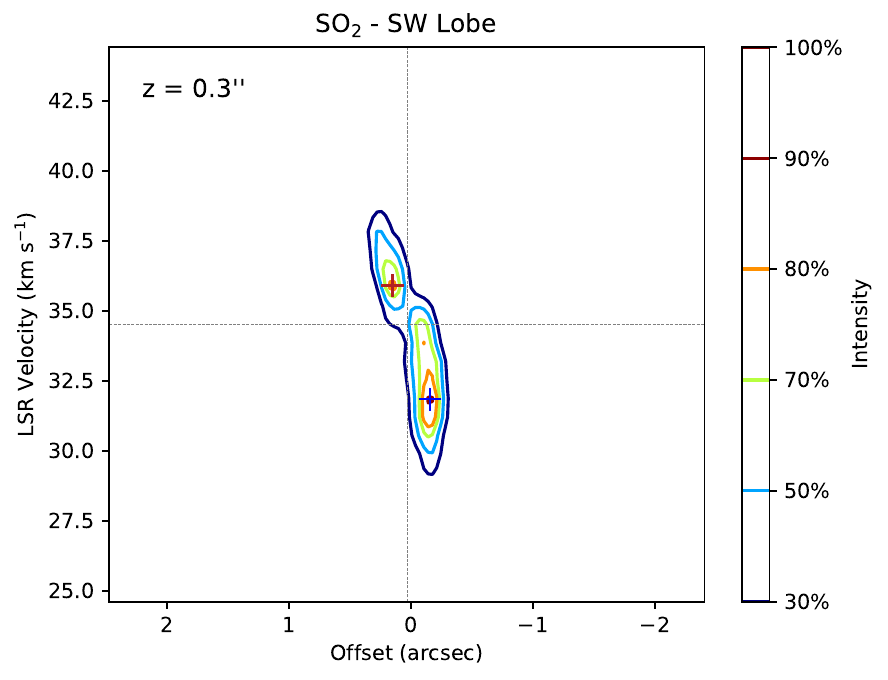}  &
  \includegraphics[width=60mm]{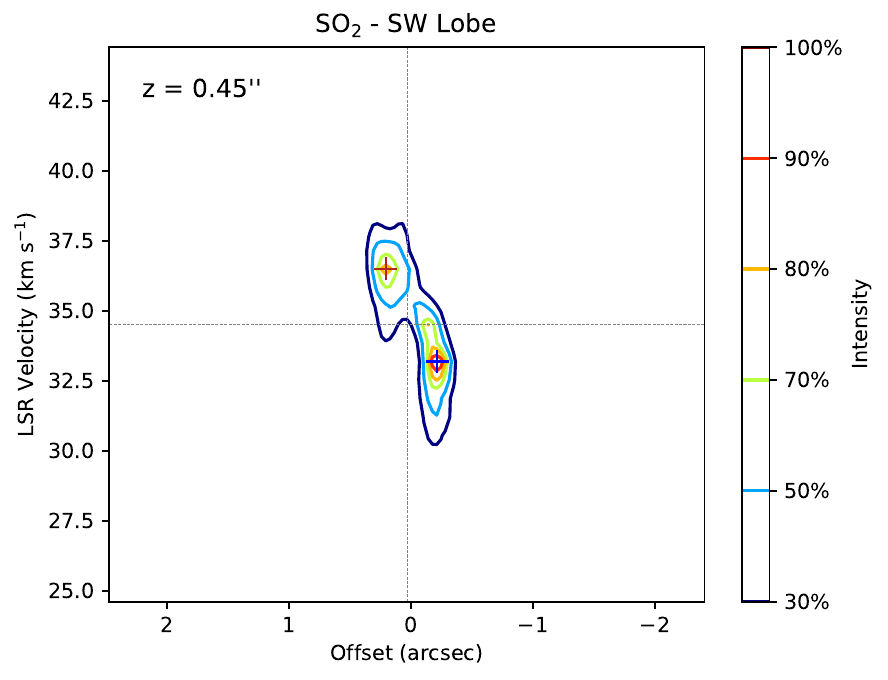}  \\
   \includegraphics[width=60mm]{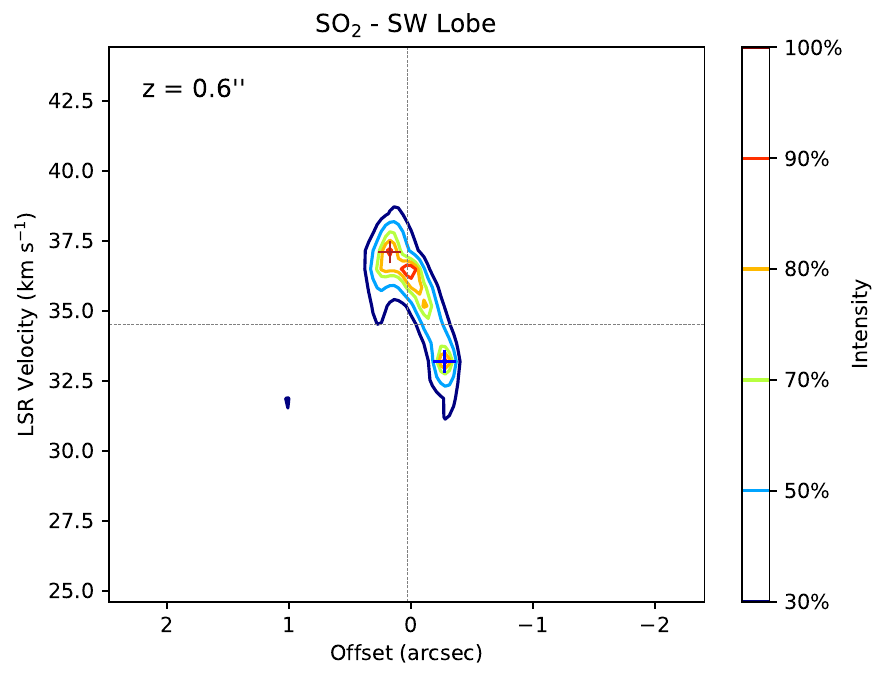}  &
  \includegraphics[width=60mm]{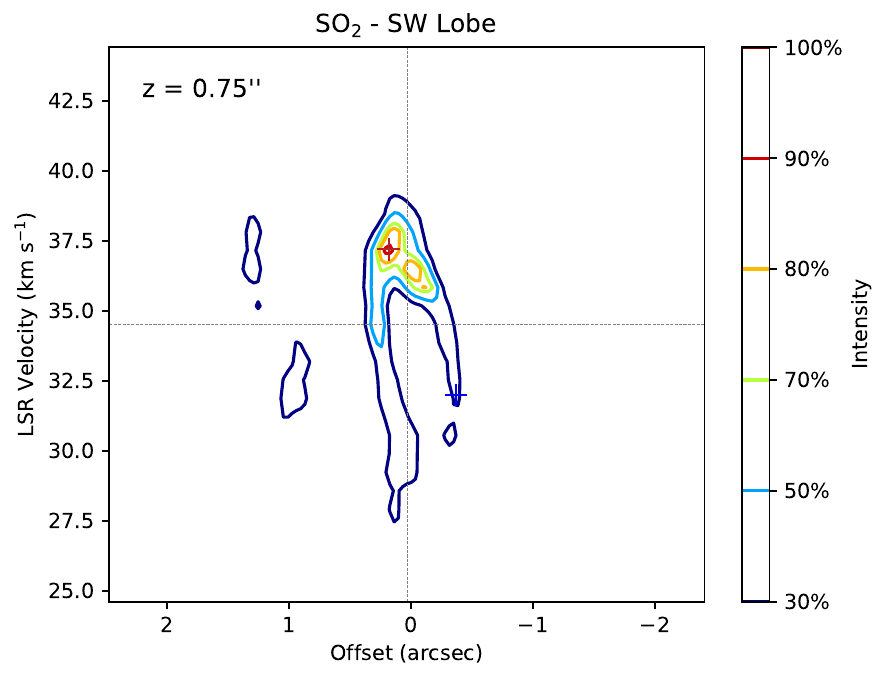}  &
\end{tabular}
\caption{Same as Fig. A.2 but for the SW lobe.}
\label{fig:pvso2sw}
\end{figure*}

\end{appendix}
\end{document}